\documentclass[aip,jcp,amsfonts,amssymb,amsmath,reprint,superscriptaddress,floatfix,longbibliography]{revtex4-2}
\usepackage{graphicx}
\usepackage{bm}
\usepackage{multirow}
\usepackage{color}
\usepackage{braket}
\usepackage{algorithm}
\usepackage{algpseudocode}
\usepackage{setspace}
\usepackage{subcaption}
\usepackage[colorlinks]{hyperref}
\hypersetup{ 
    colorlinks=true,
    linkcolor=red!80!black,
    urlcolor=magenta!80!black,
    citecolor=green!70!black,
}
\usepackage{upgreek}
\usepackage{booktabs}
\usepackage[round-mode=places,round-precision=6]{siunitx}
 \usepackage[T1]{fontenc}
\usepackage{dcolumn}
\usepackage{xcolor}
\usepackage{verbatim}
\usepackage{caption}
\usepackage{threeparttable}
\newcommand*{\citen}[1]{%
  \begingroup
    \romannumeral-`\x 
    \setcitestyle{numbers}%
    \cite{#1}%
  \endgroup
}
\makeatletter
\newcommand*{\rom}[1]{\expandafter\@slowromancap\romannumeral #1@}
\makeatother

\newcommand{\ee}{\text{e}}
\newcommand{\ii}{\text{i}}

\newcommand{\bs}[1]{\boldsymbol{ #1 }}

\begin{document}

\title{Magnetic Optical Rotation from Real-Time Simulations in Finite Magnetic Fields}

\author{Benedicte Sverdrup Ofstad}
\email{b.s.ofstad@kjemi.uio.no}
\affiliation{Hylleraas Centre for Quantum Molecular Sciences, Department of Chemistry, University of Oslo, Norway}
\author{Meilani Wibowo-Teale}
\affiliation{School of Chemistry, University of Nottingham, University Park, Nottingham NG7 2RD, United Kingdom}  
\author{H{\aa}kon Emil Kristiansen}
\affiliation{Hylleraas Centre for Quantum Molecular Sciences, Department of Chemistry, University of Oslo, Norway}
\author{Einar Aurbakken}
\affiliation{Hylleraas Centre for Quantum Molecular Sciences, Department of Chemistry, University of Oslo, Norway}
\author{Marios Petros Kitsaras}
\affiliation{University of Saarland, Physical and Theoretical Chemistry, Campus B2.2, 66123, Saarbruecken, Germany}
\author{{\O}yvind Sigmundson Sch{\o}yen}
\affiliation{Department of Physics, University of Oslo, Norway}
\author{Eirill Hauge}
\affiliation{Hylleraas Centre for Quantum Molecular Sciences, Department of Chemistry, University of Oslo, Norway}
\affiliation{Department of Numerical Analysis and Scientific Computing, Simula Research Laboratory, 0164 Oslo, Norway}
\author{Tom J. P. Irons}
\affiliation{School of Chemistry, University of Nottingham, University Park, Nottingham NG7 2RD, United Kingdom}      
\author{Simen Kvaal}
\affiliation{Hylleraas Centre for Quantum Molecular Sciences, Department of Chemistry, University of Oslo, Norway}        
\author{Stella Stopkowicz}
\affiliation{Hylleraas Centre for Quantum Molecular Sciences, Department of Chemistry, University of Oslo, Norway}
\affiliation{University of Saarland, Physical and Theoretical Chemistry, Campus B2.2, 66123, Saarbruecken, Germany}
\author{Andrew M. Wibowo-Teale}
\affiliation{Hylleraas Centre for Quantum Molecular Sciences, Department of Chemistry, University of Oslo, Norway}
\affiliation{School of Chemistry, University of Nottingham, University Park, Nottingham NG7 2RD, United Kingdom}    
\author{Thomas Bondo Pedersen}
\email{t.b.pedersen@kjemi.uio.no}
\affiliation{Hylleraas Centre for Quantum Molecular Sciences, Department of Chemistry, University of Oslo, Norway}

\date{\today}

\begin{abstract}
We present a numerical approach to magnetic optical rotation based on real-time time-dependent electronic-structure theory. Not
relying on perturbation expansions in the magnetic-field strength, the formulation allows us to test the range of validity of the linear relation
between the rotation angle per unit path length and the magnetic-field strength that was established empirically by Verdet $160$ years ago.
Results obtained from time-dependent coupled-cluster and time-dependent current density-functional theory
are presented for the closed-shell molecules H$_2$, HF, and CO in magnetic fields up to $55\,\text{kT}$ at standard
temperature and pressure conditions.
We find that Verdet's linearity remains valid up to roughly $10$--$20\,\text{kT}$,
above which significant deviations from linearity are observed.
Among the three current density-functional approximations tested in this work,
the current-dependent Tao-Perdew-Staroverov–Scuseria hybrid functional performs the best in
comparison with time-dependent coupled-cluster singles and doubles results for the magnetic optical
rotation.
\end{abstract}

\maketitle

\section{Introduction}\label{sec:intro}

Nonperturbative approaches to matter-field interactions have found increasing use in electronic-structure theory over the past decades.
Not only do such approaches allow one to overcome the inherent limitations of perturbation theory---weak field strengths, adiabatic
switching-on of the fields, (lack of) convergence of the perturbation series---they
also pave the way for fundamental discoveries, such as
the paramagnetic bonding mechanism in strong magnetic fields~\cite{lange_paramagnetic_2012} or the quantum-dynamical 
mechanism underpinning high-harmonic generation.~\cite{Lewenstein1994}

Using atom-centered Gaussian basis sets, quantum-chemical studies of electronic ground and bound excited states
in strong magnetic fields up to about
one atomic unit ($1\,\text{B}_0 = m_\text{e}E_\text{h}/\text{e}\hbar \approx 235\,052\,\text{T}$) are often motivated
by the astrophysical search for heavier atoms and even complex polyatomic molecules in, e.g., the atmosphere of magnetic white
dwarfs.~\cite{hollands_dz_2023} Such calculations require some modifications of existing implementations of commonly used
electronic-structure theories such as Hartree--Fock theory,~\cite{Tellgren2008,sen_excited_2019,sun_ab_2019,sun_generalized_2019}
full configuration-interaction (FCI) theory,~\cite{lange_paramagnetic_2012}
and coupled-cluster (CC) theory.~\cite{stopkowicz_coupled-cluster_2015,hampe_equation--motion_2017,hampe_transition-dipole_2019,hampe_full_2020}
In the case of density-functional theory (DFT), however, a more fundamental change is required since the density-functional approximation
must depend on the current density in addition to the electron density to be applicable to electrons in a static uniform magnetic
field.~\cite{Vignale1987,Vignale1988,Tellgren2012b,Tellgren2014a,Furness2015,Kvaal2021}
As an added benefit, the resulting current density-functional theory (CDFT)~\cite{Tellgren2014a} is applicable also at low
magnetic-field strengths, thus potentially improving upon results obtained from conventional DFT for,
e.g., nuclear magnetic resonance shielding constants.

Similarly, real-time time-dependent electronic-structure theory~\cite{goings_real-time_2018,li_real-time_2020}
allows one to simulate laser-induced quantum dynamics without invoking perturbation theory.
This is particularly important for the simulation of time-resolved spectroscopies where a pump laser
is used to create a non-stationary electronic wave packet, which is then probed by a second laser pulse
applied at variable delays relative to the pump laser. With attosecond laser pulses, it thus becomes possible
to observe and manipulate ultrafast electronic motion,
opening a new field of chemistry---attochemistry.~\cite{nisoli_attosecond_2017,palacios_quantum_2020}
Moreover, by carefully choosing laser-pulse parameters such as shape, electric-field strength, and carrier frequency,
it becomes possible to extract complete linear absorption spectra, including bound core as well as valence excitations,
and low-order nonlinear optical properties from a single or a few simulations.~\cite{li_real-time_2020}

By combining the nonperturbative treatment of static uniform magnetic fields with real-time time-dependent electronic-structure
theory of laser-driven multi-electron dynamics we get access to magnetic-field-induced, frequency-dependent
molecular properties such as magnetic optical rotation (MOR).
Also named the Faraday effect after its discoverer,~\cite{faraday_i_1846}
MOR is the rotation of the polarization plane of linearly polarized light passing through a transparent medium
in the presence of a magnetic field with nonzero component along the propagation direction of the light beam.
While phenomenologically similar,~\cite{buckingham_magnetic_1966} 
MOR differs from natural optical rotation by being observable for chiral and achiral molecules alike.

The Faraday effect, which was of pivotal importance for the development
of Maxwell's electromagnetic theory,~\cite{knudsen_faraday_1976}
is today exploited in a large number of technological applications, including fiber-optic current sensors,~\cite{Hui_fiber-based_2023}
satellite communication systems,~\cite{meyers_communications_2003} and
measurement of interstellar and intergalactic magnetic fields.~\cite{han_observing_2017}
Consequently, much research effort remains invested in the development of high-MOR materials.~\cite{carothers_high_2022}
The foundation of all these applications is the linear relationship,
discovered through a series of thorough experiments by Verdet,~\cite{verdet,verdet2}
between the rotation angle $\theta$ per unit path length $l$ and the magnetic-field strength $B$ along
the propagation direction of the light: $\theta/l = VB$. The constant of proportionality, the Verdet constant $V$,
was found by Verdet to depend on the frequency of the light and to be characteristic of each type of
molecule such that $V$ for a given solution can be obtained by summation over the contributions from each 
solute and solvent molecule.~\cite{verdet,verdet2,verdet3,verdet4}
This observation implies that a quantum-mechanical account of the microscopic origin of MOR can be approximately reduced to
the response of a single molecule to a static uniform magnetic field and a time-dependent uniform electric
field.~\cite{Serber1932,buckingham_magnetic_1966,Barron2004}
Accordingly,
\citeauthor{Parkinson1997} formulated the Verdet constant in terms of a mixed magnetoelectric quadratic response
function, i.e., a third-order mixed perturbation theory, linear in the magnetic field and quadratic in the
electric field.~\cite{Parkinson1997}
The response-function approach has been used to compute Verdet constants at the CC level of theory
by Coriani et al.~\cite{coriani_coupled_1997,coriani_gauge-origin_2000,coriani_triple_2000}

Avoiding the perturbation expansion in the magnetic-field strength,
we will in this work investigate the range of validity of the linear relationship between $\theta/l$ and $B$
using real-time time-dependent coupled-cluster (TDCC) theory~\cite{review2023}
and real-time time-dependent current density-functional theory (TDCDFT).~\cite{wibowo_modeling_2021}
This allows, for the first time, a direct comparison of electron dynamics simulations at the
TDCDFT and TDCC levels of theory.

This paper is organized as follows.
The general quantum-mechanical theory of MOR is outlined in Sec.~\ref{sec:theory} along with a description of how
real-time time-dependent electronic-structure simulations may be utilized to compute $\theta/l$ without perturbation expansion
in $B$. In Sec.~\ref{sec:results} we validate our magnetic-field-dependent implementation of various TDCC theories
by comparing linear absorption spectra from simulations at finite $B$ with those obtained from equation-of-motion
excitation-energy coupled-cluster
(EOM-EE-CC) theory,~\cite{hampe_equation--motion_2017} followed by presentation and discussion of MOR results obtained
from TDCDFT and TDCC simulations. Finally, Sec.~\ref{sec:conc} contains our concluding remarks.

\section{Theory}\label{sec:theory}

\subsection{Magnetic optical rotation}

The polarization plane of linearly polarized light propagating parallel to a static uniform magnetic field through a sample
of molecules is rotated by an angle $\theta_{r}$, given by~\cite{Serber1932,buckingham_magnetic_1966,Barron2004}
\begin{equation}\label{free_mor}
    \frac{\theta_{r}}{\ell} = \frac{1}{3} C \omega \sum_{ijk} \epsilon_{ijk} \text{Im}\left[\alpha^{(k)}_{ij}(-\omega;\omega)\right],
\end{equation}
where $\ell$ is the path length, $\omega$ is the angular frequency of the light, $\alpha_{ij}^{(k)}(-\omega;\omega)$
is the Cartesian $ij$ component of the molecular polarizability tensor in the presence of the magnetic field along axis $k$,
and $\epsilon_{ijk}$ denotes the Levi--Civita symbol. The subscript $r$ signifies that the sample of molecules are randomly oriented.
The constant $C$ is given by
\begin{equation}
\label{constant}
	C =  \frac{1}{2c} \left( \frac{2 \pi \mathcal{N}}{4 \pi \epsilon_0} \right),
\end{equation} 
where $\mathcal{N}$ is the number density, $c$ is the speed of light, and $\epsilon_0$ is the vacuum permittivity. 

For weak magnetic fields, the polarizability $\alpha^k_{ij}$ may be expanded to first order in the magnetic-field strength $B$, leading to
the conventional formula
\begin{equation}
    \label{eq:verdet}
    \frac{\theta_{r}}{\ell} = V(\omega) B,
\end{equation}
where $V(\omega)$ is the Verdet constant, which can be expressed in terms of frequency-dependent quadratic response
functions.~\cite{parkinson_calculation_1993,Parkinson1997,coriani_coupled_1997,coriani_gauge-origin_2000,coriani_triple_2000}
Avoiding the expansion in $B$, we instead work treat the magnetic field nonperturbatively and thus enable the calculation
of MOR at any magnetic-field strength.

As the magnetic-field strength grows, however, the orienting effect of the magnetic field becomes increasingly
important.~\cite{irons_optimizing_2021}
In lieu of a rigorous averaging procedure, we also compute the MOR assuming that all molecules of the sample are found in the energetically
most favorable geometry and orientation relative to the magnetic field. For the case of fixed orientation, denoted with the subscript $o$, we use the following expression
for the MOR per unit path length,
\begin{equation}\label{fixed_mor}
    \frac{\theta_{o}}{\ell} = C \omega \text{Im}\left[\alpha^{(z)}_{xy}(-\omega;\omega)\right],
\end{equation}
where the magnetic field vector is chosen parallel to the $z$-axis.

We now turn to the problem of extracting the complex frequency-dependent polarizability from simulations of laser-driven
electron dynamics in the presence of a static uniform magnetic field.

\subsection{Electron dynamics in a finite magnetic field}

We consider a nonrelativistic atomic or molecular system with $N$ electrons exposed to a static uniform magnetic field $\bs{B}$
and a time-dependent radiation field with the electric and magnetic fields $\bs{\mathcal{E}}(\bs{r},t)$ and $\bs{\mathcal{B}}(\bs{r},t)$.
Although vibrational effects must be taken into account for highly accurate calculations of the MOR,\cite{bishop_magnetic_1990, parkinson_calculation_1993, coriani_coupled_1997}
we will only consider electronic contributions in this work.
Within the clamped-nuclei Born-Oppenheimer approximation, the electronic minimal-coupling Hamiltonian can be written
as (we use atomic units throughout)
\begin{align}\label{eq:mcH}
    \hat{H}(t) &= \sum_{i=1}^N \left\{ \frac{1}{2}\hat{\pi}_i^2(\bs{r}_i,t) + \hat{\bs{S}}_i \cdot [\bs{B} + \bs{\mathcal{B}}(\bs{r}_i,t)] \right\}
    \nonumber \\
    &+ W_\text{en} + W_\text{ee},
\end{align}
where $W_\text{en}$ is the electronic-nuclear Coulomb attraction, $W_\text{ee}$ is the electronic-electronic Coulomb repulsion, and
$\hat{\bs{S}}_i$ and $\bs{r}_i$ are the spin and position operators, respective, of electron $i$. The
constant nuclear repulsion energy is excluded for convenience and the kinetic momentum operator,
\begin{equation}\label{eq:kinetic_momentum}
    \hat{\bs{\pi}}(\bs{r},t) = \hat{\bs{p}} + \bs{A}(\bs{r}) + \bs{\mathcal{A}}(\bs{r},t),
\end{equation}
differs from the canonical momentum operator $\hat{\bs{p}} = -\ii \bs{\nabla}$ by including the time-dependent electromagnetic vector potential
$\bs{\mathcal{A}}(\bs{r},t)$ and the static magnetic vector potential
\begin{equation}\label{eq:magneticA}
    \bs{A}(\bs{r}) = \frac{1}{2} \bs{B} \times (\bs{r} - \bs{O}),
\end{equation}
where $\bs{O}$ is the magnetic gauge origin.
The Coulomb gauge condition is chosen for the electromagnetic vector potential and the scalar potential vanishes
such that the source-free electric and magnetic fields are given by
\begin{equation}
    \bs{\mathcal{E}}(\bs{r},t) = -\partial_t \bs{\mathcal{A}}(\bs{r},t), \qquad
    \bs{\mathcal{B}}(\bs{r},t) = \bs{\nabla} \times \bs{\mathcal{A}}(\bs{r},t).
\end{equation}

The Hamiltonian can be recast as
\begin{equation}
    \hat{H}(t) = \hat{H}_0 + \hat{V}(t),
\end{equation}
where the time-independent Hamiltonian
\begin{equation}\label{eq:magnetic_hamiltonian}
    \hat{H}_0 = \sum_{i=1}^N \left( \frac{1}{2}(\hat{\bs{p}}_i+ \bs{A}(\bs{r}_i))^2 + \hat{\bs{S}}_i \cdot \bs{B} \right)
    + W_\text{en} + W_\text{ee},
\end{equation}
describes the electronic system interacting with the static uniform magnetic field $\bs{B}$,
and
\begin{align}\label{eq:full_V}
    \hat{V}(t) = \sum_{i=1}^N \Big(&
        \left[ \hat{\bs{p}}_i + \bs{A}(\bs{r}_i) \right] \cdot \bs{\mathcal{A}}(\bs{r}_i,t)
        +  \hat{\bs{S}}_i \cdot \bs{\mathcal{B}}(\bs{r}_i,t) \nonumber \\
        &+ \frac{1}{2} \mathcal{A}^2(\bs{r}_i,t)
    \Big),
\end{align}
describes the interaction with the time-dependent radiation field.
Unless $\bs{B}$ and $\bs{E}(\bs{r},t)$ are parallel,
the dependence on the static uniform magnetic field cannot be isolated in the energy operator $\hat{H}_0$
but appears in the time-dependent interaction operator as well.

The time-dependent Schr{\"o}dinger equation reads
\begin{equation}\label{eq:TDSE}
    \ii \partial_t \Psi(t) = \hat{H}(t) \Psi(t), \qquad \Psi(0) = \Psi_0,
\end{equation}
where $\Psi_0$ is the wave function of the initial electronic state before the radiation field is switched on. We choose $\Psi_0$ 
to be the magnetic field-dependent ground-state wave function $\psi_0$ with energy $E_0$,
\begin{equation}\label{eq:TISE}
    \hat{H}_0 \psi_0 = E_0 \psi_0.
\end{equation}
This equation can be solved approximately using recent implementations of quantum-chemical methods such as
Hartree--Fock theory\cite{Tellgren2008},
coupled-cluster theory,~\cite{stopkowicz_coupled-cluster_2015}
and current density-functional theory.~\cite{Tellgren2014a}
With finite-dimensional, isotropic Gaussian-type orbital basis sets,
magnetic gauge-origin invariance can be maintained by multiplying each basis function
with a magnetic field-dependent phase factor to obtain the so-called
London atomic orbitals (LAOs).~\cite{lao}
This approach works well for ground and excited states in magnetic fields up to about $1\,\text{B}_0$,
whereas anisotropic Gaussians or high angular momenta are required for even stronger magnetic fields.~\cite{Schmelcher1988,lehtola_fully_2020}

The interaction operator of Eq.~\eqref{eq:full_V} can be substantially simplified by assuming the electric-dipole approximation,
$\bs{\mathcal{A}}(\bs{r},t) \approx \bs{\mathcal{A}}(\bs{0},t) \equiv \bs{\mathcal{A}}(t)$, which is valid for radiation wavelengths
well beyond the ``size'' of the atomic or molecular system. In the context of MOR, we are interested in the transparent spectral region
of the molecules studied, i.e., in energies below the first excitation energy, and the conditions for using the electric-dipole
approximation thus are satisfied.
A simple gauge transformation then yields the usual length-gauge dipole interaction
operator,
\begin{equation}\label{eq:interaction_operator}
    \hat{V}(t) = \sum_{i=1}^N \hat{\bs{r}}_i \cdot \bs{\mathcal{E}}(t), 
\end{equation} 
where $\bs{\mathcal{E}}(t) = -\partial_t \bs{\mathcal{A}}(t)$.
Thus, within the electric-dipole approximation, the interaction operator is independent of the static uniform magnetic field.
The electric-dipole interaction operator is assumed in the following sections where we briefly summarize
the TDCC and TDCDFT approaches to the simulation of laser-driven electron dynamics in a static uniform magnetic field.

\subsection{Time-dependent electronic-structure theory}

\subsubsection{Time-dependent coupled-cluster theory}

Providing systematically improvable ground- and excited-state energies and properties, the CC hierarchy of wave-function
approximations~\cite{MEST,Crawford2000,Bartlett2007,Shavitt2009,Krylov2008,Sneskov2012,Helgaker2012}
is arguably the most successful wave function-based approach to the calculation of atomic and molecular electronic structure. 
At least for systems with a nondegenerate ground state dominated by a single Slater determinant, CC calculations---especially with 
the ``Gold Standard'' of quantum chemistry, the CC singles and doubles with perturbative triples (CCSD(T))~\cite{Raghavachari1989}
model---are generally more reliable than and can serve as benchmarks for affordable density-functional approximations within DFT.

In the past decade, CC theory for ground~\cite{stopkowicz_coupled-cluster_2015,hampe_full_2020} and
excited~\cite{hampe_full_2020,hampe_equation--motion_2017} states has been developed for molecules in finite magnetic fields, i.e.,
for Hamiltonians of the form given in Eq.~\eqref{eq:magnetic_hamiltonian}. Compared with a conventional field-free implementation of
CC theory,~\cite{MEST,Crawford2000} the main complications arising from the finite magnetic field are that the molecular orbitals
and cluster amplitudes
necessarily become complex and that the dependence on the magnetic gauge-origin must be eliminated. As mentioned above,
the latter can be elegantly handled by using LAO basis functions~\cite{Tellgren2008} and suitably modified integral-evaluation
algorithms.~\cite{Tellgren2008,Tellgren2012a,irons_efficient_2017}
The complex orbitals, however, reduce the permutational symmetries of
the one- and two-electron integrals. As long as these permutational-symmetry reductions are properly taken into account,
a conventional CC implementation can be rather straightforwardly turned into a magnetic-field implementation by switching
from real to complex arithmetic.
Similarly, an implementation of TDCC theory~\cite{review2023} only requires complex arithmetic in the ground-state
CC functions, and may thus be used without modifications to simulate electron dynamics in finite magnetic fields provided that the
reduced integral permutation symmetry is properly handled.

In this work we use two different classes of TDCC theory. In the first class, the single reference determinant is chosen to be
the Hartree--Fock ground-state Slater determinant obtained from the magnetic Hamiltonian given in Eq.~\eqref{eq:magnetic_hamiltonian}. The TDCC singles-and-doubles (TDCCSD)~\cite{pedersen_symplectic_2019} model and its second-order approximation,
the TDCC2 model,~\cite{Christiansen1995,omp2} belong to this class.
In the second class, the single reference determinant is built from time-dependent spin orbitals which are 
bivariationally optimized alongside the cluster amplitudes.
The orbital relaxation is constrained to conserve orthonormality in 
orbital-optimized TDCC (TDOCC) theory,~\cite{pedersen_gauge_1999,sato_communication_2018}
whereas nonorthogonal orbital-optimized TDCC (TDNOCC) theory~\cite{pedersen_gauge_2001,kvaal_ab_2012}
requires biorthonormal orbitals. In both cases, the orbital relaxation makes
the singles cluster operators redundant.~\cite{pedersen_gauge_1999,sato_communication_2018,pedersen_gauge_2001,kvaal_ab_2012}
Here we use the TDNOCCD model,~\cite{pedersen_gauge_2001,kristiansen_numerical_2020} which includes doubles amplitudes only, and
the time-dependent orbital-optimized second-order M{\o}ller-Plesset (TDOMP2) method,~\cite{omp2,Pathak2020} which is a second-order
approximation analogous to TDCC2 theory. The TDCCSD and TDNOCCD methods exhibit a computational scaling of
$\mathcal{O}(K^6)$ with respect to the number of basis functions $K$, while the TDCC2 and TDOMP2 models scale as $\mathcal{O}(K^5)$.

\subsubsection{Time-dependent current density functional theory}

The density-functional theory (DFT) is extended to take into account the effect of an external magnetic field by including a dependence on both the charge density $\rho$ and the paramagnetic component of the induced current density $\mathbf{j}_{\mathrm{p}}$ in the universal density functional $F[\rho, \mathbf{j}_{\mathrm{p}}]$. It was shown in Refs.~\citen{Tellgren2012b,Kvaal2021} that the Vignale--Rasolt formulation~\cite{Vignale1987,Vignale1988} of current-DFT (CDFT) can be treated in a similar manner to Lieb's formulation~\cite{Lieb1983} of conventional DFT.

A non-perturbative treatment of an external magnetic field in the Kohn--Sham CDFT scheme can be set up by using LAOs (see Refs.~\citen{Vignale1988,Tellgren2014a,Furness2015} for details of the resulting Kohn-Sham equations). A central challenge in CDFT calculations is then to define the exchange--correlation functional $E_{\mathrm{xc}}[\rho, \mathbf{j}_{\mathrm{p}}]$, which now also depends on both the charge- and paramagnetic current densities. It has been shown that the accuracy of CDFT calculations using vorticity-based corrections to local density approximation (LDA) and generalised gradient approximation (GGA) levels is poor~\cite{Lee1995,Zhu2006,Tellgren2014a}. However, introducing an explicit current dependence at the meta-GGA level via a modification of the kinetic energy density as suggested by \citeauthor{Dobson1993}~\cite{Dobson1993} and later used by \citeauthor{Becke1996}~\cite{Becke1996} and \citeauthor{Bates2012}~\cite{Bates2012} as
\begin{equation}
	\tau(\mathbf{r}) \rightarrow \tilde{\tau}(\mathbf{r}) =
    \sum_{i}^{\mathrm{occ}} [\nabla \varphi_i(\mathbf{r})]^{\ast} \cdot
    [\nabla \varphi_{i}(\mathbf{r})] - \frac{\lvert\mathbf{j}_{\mathrm{p}}(\mathbf{r})\rvert^2}{\rho(\mathbf{r})},
\end{equation}
leads to a well-defined and properly bounded iso-orbital indicator when it is applied to the Tao-Perdew-Staroverov–Scuseria (TPSS) functional~\cite{Scuseria2003}, which in this work is denoted as cTPSS. Two variants of hybrid cTPSS functionals, denoted as cTPSSh and cTPSSrsh, have been introduced to dynamic CDFT in Ref.~[\citen{wibowo_modeling_2021}].

The cTPSSh functional includes a mixture of $10 \%$ orbital-dependent exchange and $90 \%$ cTPSS exchange functionals for the exchange contribution and $100 \%$ of the cTPSS correlation contribution. The cTPSSrsh functional similarly consists of a $100 \%$ cTPSS correlation functional, and a range-separated TPSS-like exchange functional defined in Ref.~[\citen{irons_analyzing_2020}].
In this work we will use the cTPSS, cTPSSh, and cTPSSrsh functionals for computing the magnetic optical activity and compare their performance relative to the TDCC models above.

\subsection{Extracting dynamic properties}\label{subsec:extract}

Depending on the shape of the time-dependent electric field, the induced dipole moment computed
as an expectation value of the electric-dipole operator at each time step during a simulation
can be used to extract properties such as absorption spectra and polarizabilities. We first consider the generation of
linear absorption spectra, which we will later use to validate the TDCC implementations by comparison with spectra
computed through time-independent
EOM-EE-CC~\cite{Stanton1993,Krylov2008,hampe_equation--motion_2017} theory.
We then describe the extraction procedure used to obtain complex
polarizabilities, which are subsequently combined to yield the MOR.
For notational convenience, we do not use the superscript $k$ to denote the direction of the
magnetic field in this section.

\subsubsection{Absorption spectra}

In order to extract excitation energies and intensities using time-dependent electronic-structure theory,
the external electric field $\bs{\mathcal{E}}(t)$ in Eq.~\eqref{eq:interaction_operator} is chosen as a
$\delta$-pulse at $t=0$, $\bs{\mathcal{E}}(t) = \mathcal{E}\bs{u}\delta(t)$ where $\mathcal{E}$ is the field strength,
$\bs{u}$ is the (linear) unit polarization vector, and $\delta(t)$ is the Dirac delta function. Completely localized in time, this pulse is infinitely broad in the frequency domain
and, therefore, excites the electronic system from the ground state into all electric-dipole allowed excited states.
If the field strength is sufficiently weak, nonlinear processes such as multiphoton transitions and transitions among
excited states are negligible, resulting in a linear absorption spectrum.

The $\delta$-pulse is approximated as a box function,
\begin{equation}\label{eq:delta_pulse}
    \bs{\mathcal{E}}(t) =
	\begin{cases}
        \mathcal{E}\bs{u} & 0 \leq t  \leq \Delta t, \\
		0 & \qquad \text{else},
	\end{cases}
\end{equation}
where $\Delta t$ is the time step of the simulation---i.e., the field is on during the first time step only.
The absorption spectrum is given by
\begin{equation}\label{eq:intensity}
    S(\omega) = \frac{4\pi \omega}{3c} \text{Im} \sum_i \mu_i(\omega),
\end{equation}
where $\mu_i(\omega)$ is obtained as the discrete Fourier transform, computed with the normalized fast Fourier transform (FFT) of the time-dependent dipole moment induced along Cartesian axis $i$ by a $\delta$-pulse polarized along the same axis,
\begin{equation}\label{eq:frequency}
    \mathbf{\mu}_i(\omega) = \text{FFT}[(\mu_{i}(t) - \mu_i^{(0)}) \ee^{-\gamma t}]/\mathcal{E},
\end{equation}
where $\mu_i^{(0)}$ is the permanent ground-state dipole moment, and $\mu_i(t)$ is the dipole moment computed at time $t$.
The damping factor $\exp(-\gamma t)$ is applied to avoid artefacts from the periodic FFT procedure and the parameter
$\gamma > 0$ can be interpreted as a common (inverse) lifetime of the excited states, causing Lorentzian shapes of the
absorption lines.
In general, the calculation of the linear absorption spectrum thus requires three independent simulations, one for each Cartesian direction. To ensure satisfactory resolution in the resulting spectra, the duration of the simulations need to be relatively long, typically with total simulation times exceeding $1000$ a.u.

\subsubsection{Complex polarizabilities}

The frequency-dependent polarizability $\alpha_{ij}(-\omega;\omega)$ can be extracted from simulations
using the ramped continuous wave approach of \citeauthor{ding_efficient_2013}~\cite{ding_efficient_2013}
with the quadratic ramp proposed by \citeauthor{ofstad_adiabatic_2023}~\cite{ofstad_adiabatic_2023} to
suppress nonadiabatic effects. In the present case, however, it must be recalled that the polarizability is
complex in the presence of the magnetic field.

With linear polarization vector $\bs{u}$ and field strength $\mathcal{E}$, 
the electric field takes the form~\cite{ofstad_adiabatic_2023}
\begin{equation}\label{eq:quadratic}
    \bs{\mathcal{E}}(t) =
		\begin{cases}
            \frac{2t^2}{t_{n_r}^2} \mathcal{E}\bs{u} \sin(\omega t) &  0 \leq t < \frac{t_{n_r}}{2}, \\
            \frac{t_{n_r}^2 - 2(t-t_{n_r})^2}{t_{n_r}^2}\mathcal{E}\bs{u} \sin(\omega t)  &   \frac{t_{n_r}}{2} \leq t < t_{n_r}, \\
            \mathcal{E}\bs{u} \sin(\omega t)  &    t_{n_r} \leq t \leq t_\text{tot},
		\end{cases}
\end{equation}
where $t_{n_r}$ is the ramping time expressed as a multiple $n_r$ of optical cycles,
\begin{equation}
    t_{n_r} = n_r \frac{2\pi}{\omega},
\end{equation}
after which the electric field remains a full-strength, monochromatic continuous wave until the simulation concludes at $t=t_\text{tot}$.

For sufficiently weak electric-field strengths,
the Cartesian component $i$ of the electric dipole moment computed at times $t_{n_r} \leq t \leq t_\text{tot}$ can be expanded as
\begin{equation}
    \label{eq:dipole_expansion}
    \mu_i(t) = \mu_i^{(0)} + \sum_{j} \alpha_{ij}(t) \mathcal{E}_j + \cdots,
\end{equation}
where
\begin{align}
    \alpha_{ij}(t) &= \text{Re}\left[\alpha_{ij}(-\omega; \omega)\right]\sin(\omega t) \nonumber \\
                   &+ \text{Im}\left[\alpha_{ij}(-\omega; \omega)\right]\cos(\omega t).
\end{align}
The time-domain polarizability $\alpha_{ij}(t)$ is computed by a four-point finite difference formula, followed by
a fitting procedure to obtain
the frequency-domain polarizability $\alpha_{ij}(-\omega; \omega)$
as described in Refs.~\onlinecite{ding_efficient_2013,ofstad_adiabatic_2023}.

\begin{figure*}
\begin{center}
\begin{subfigure}{0.44\textwidth}
\includegraphics[width=\textwidth]{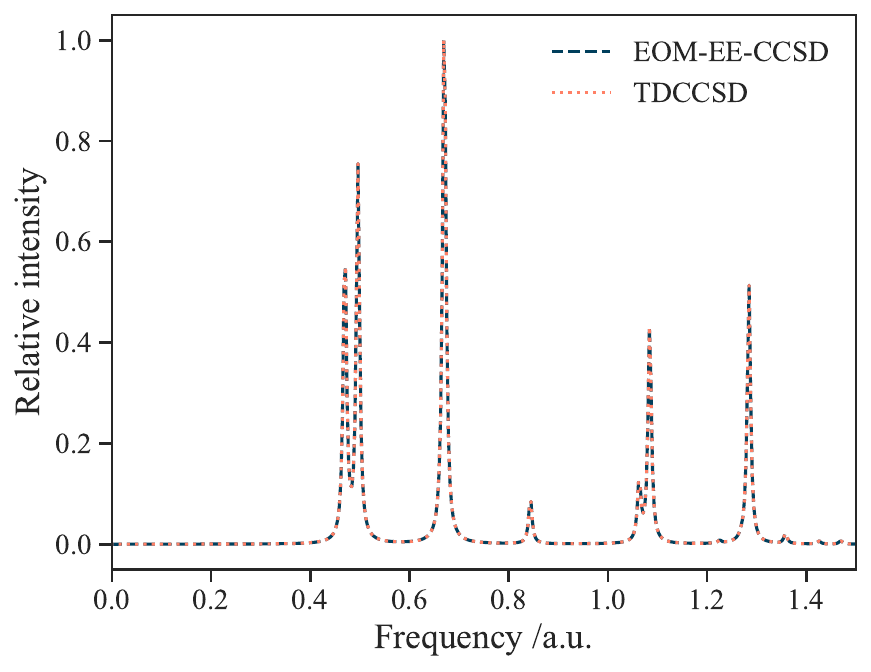} 
\caption{\centering H$_2$}
\label{abs_h2}
\end{subfigure}
\begin{subfigure}{0.44\textwidth}
\includegraphics[width=\textwidth]{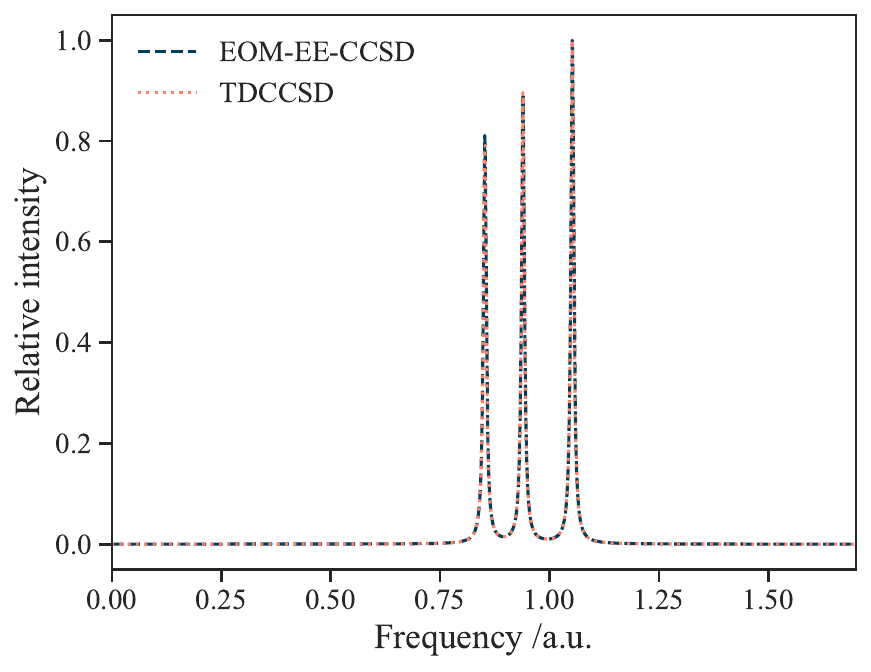} \\
\caption{\centering He}
\end{subfigure}
\begin{subfigure}{0.44\textwidth}
\includegraphics[width=\textwidth]{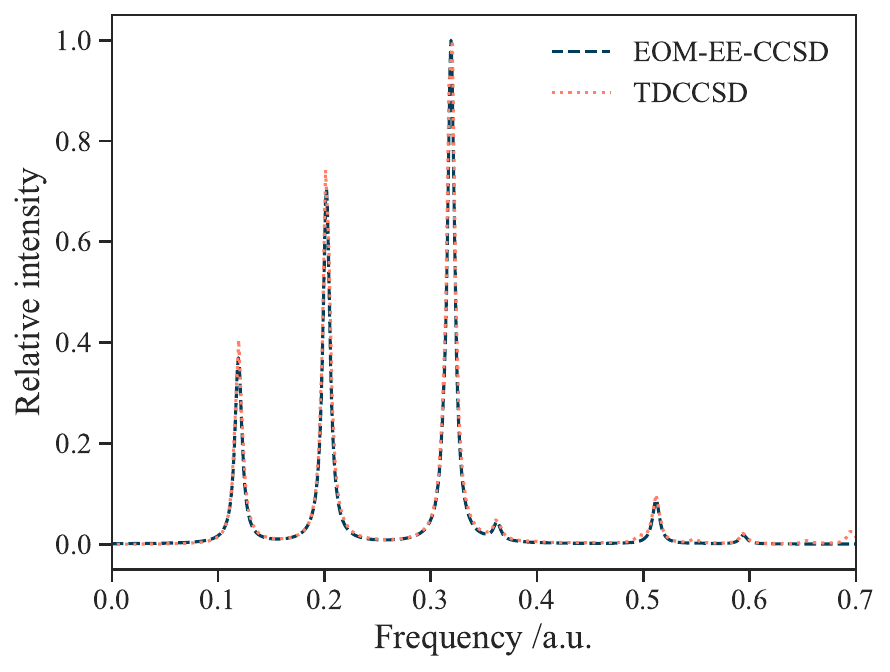} 
\caption{\centering Be}
\label{sec_hyp_thg}
\end{subfigure}
\begin{subfigure}{0.44\textwidth}
\includegraphics[width=\textwidth]{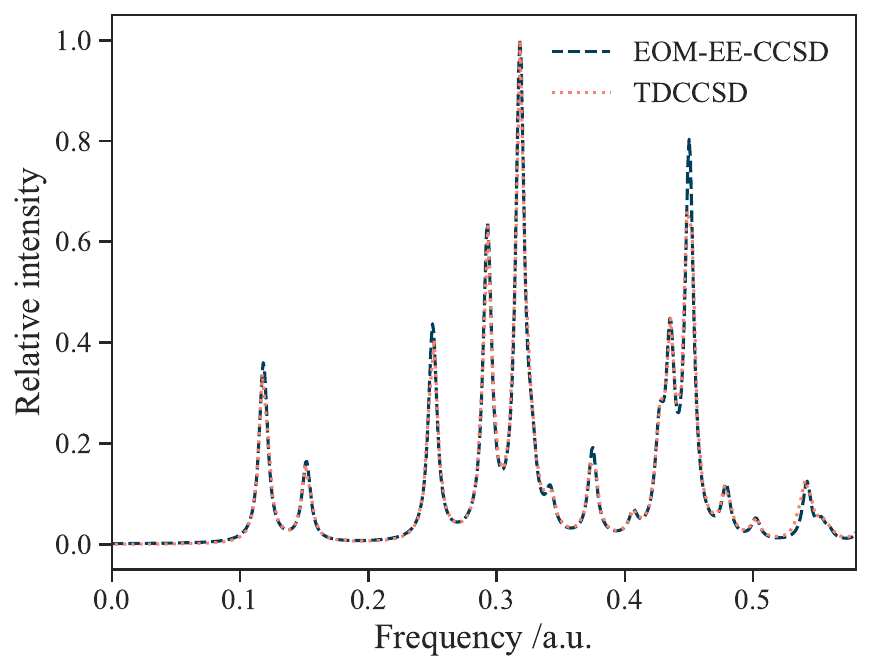} 
\caption{\centering LiH}
\end{subfigure}
    \caption{Overlay of absorption spectra generated by TDCCSD and EOM-EE-CCSD at the magnetic-field strength $0.2\,\text{B}_0$ (along the
    internuclear axis for the diatomic molecules) with the aug-cc-pVTZ basis set. \label{absorption_plots_ccsd}}
\end{center}
\end{figure*}

\section{Results}\label{sec:results}

\subsection{Absorption Spectra}\label{sec:absspec} 

We have implemented the TDCC models discussed above in the open-source \textsc{HyQD} software~\cite{HyQD} using 
the \textsc{QUEST} program~\cite{QUEST} to generate optimized Hartree--Fock orbitals and Hamiltonian integrals in LAO basis to
ensure magnetic gauge-origin invariance.~\cite{Tellgren2008}
We validate our implementation of TDCCSD theory by comparing absorption spectra obtained from $\delta$-pulse simulations
with spectra computed by the time-independent finite magnetic field EOM-EE-CCSD model~\cite{hampe_equation--motion_2017}
with the same LAO basis. The TDCCSD dipole moment is computed using the inherently real expectation-value functional
proposed in Refs.~\onlinecite{pedersen_symplectic_2019,Pedersen1997}, and the electric-field strength is
$\mathcal{E} = 0.001\,\text{a.u.}$
While the excitation energies are identical with the two approaches, the intensities may differ when
the number of electrons surpasses two. The approaches were nonetheless found to coincide quite well, with only slight deviations observed between EOM-EE-CCSD, TDCC and linear-response CC (LRCC)~\cite{hampe_transition-dipole_2019}.  Although the time-dependent approach generates the full absorption spectrum,
including core-valence excitations, we only compare low-lying transitions to avoid full diagonalization of the
EOM-EE-CC matrix.
The finite magnetic field EOM-EE-CCSD calculations are performed using the \textsc{QCUMBRE} software~\cite{QCUMBRE}
together with an interface to the \textsc{CFOUR} program package,~\cite{cfour,Matthews2020} which provides the Hartree--Fock ground-state solution using the MINT integral package.~\cite{MINT}
The EOM-EE-CCSD transition dipole moments are calculated with the expectation value approach.~\cite{Stanton1993}

Absorption spectra are computed for H$_2$, He, Be, and LiH at the magnetic-field strength $0.2\,\text{B}_0$ directed 
along the $z$-axis, which coincides with the bond axis for the diatomic molecules.
The geometries of H$_2$ and LiH are optimized in the magnetic field at the cTPSS level of theory with the
aug-cc-pVDZ~\cite{dunning_gaussian_1989,kendall_electron_1992,woon1994,prascher2011}
basis set using \textsc{QUEST}.~\cite{QUEST,irons_optimizing_2021}
The resulting bond lengths are $1.39106\,\text{a}_0$ for H$_2$ and $2.96117\,\text{a}_0$ for LiH. 

For the EOM-EE-CCSD calculations, 
a convergence threshold of $10^{-7}$ is used for the Hartree--Fock densities and CC amplitudes,
while a looser threshold of $10^{-6}$ ($10^{-5}$) is used for the right-hand (left-hand) side EOM vectors.
For the TDCCSD simulations, a convergence threshold of $10^{-12}$ for the energy-gradient norm is used for the Hartree--Fock ground state optimization,
while the CCSD ground-state amplitudes are converged to a residual norm of $10^{-12}$.
The TDCCSD equations of motion are integrated using sixth order (three-stage, $s=3$) symplectic Gauss-Legendre
integrator~\cite{integrator,pedersen_symplectic_2019} with time step $\Delta t = 0.01\,\text{a.u.}$ and residual norm
convergence criterion $10^{-10}$ for the implicit equations. The total simulation time is $1500 \,\text{a.u.}$ for He and H$_2$, and $2000\,\text{a.u.}$ for Be and LiH.

Figure \ref{absorption_plots_ccsd} displays the absorption spectra of the four systems obtained from the TDCCSD and EOM-EE-CCSD approaches
with the aug-cc-pVTZ~\cite{dunning_gaussian_1989,kendall_electron_1992,woon1994,prascher2011} basis set.
The excitation energies are equivalent to within the resolution of the FFT (approximately $0.004\,\text{a.u.}$ for He and H$_2$, and $0.003\,\text{a.u.}$ for Be and LiH) for all systems. For the two-electron systems, the intensities also agree, while minor deviations are observed for Be and LiH, as expected. 

Corresponding absorption plots obtained from TDCC2 and EOM-EE-CC2/ EOM-EE-CCSD approaches can be found in the Supplementary material.
No implementations of excitation energies for the OMP2 and NOCCD methods are available (only a pilot implementation of linear response
theory was reported in Ref.~\citenum{pedersen_gauge_2001}), and, hence, no validation is possible for these methods.

\subsection{Magnetic Optical Activity}\label{sec:magopt}

\subsubsection{Computational details}

\begin{figure}
\begin{center}
\includegraphics[scale=0.5]{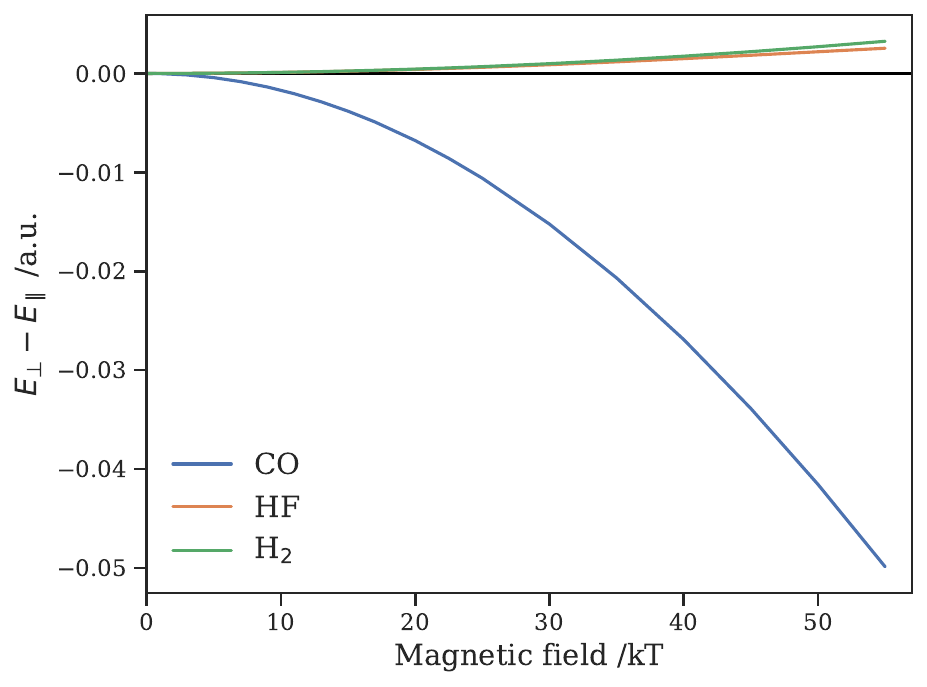} 
    \caption{The Hartree--Fock/aug-cc-pVDZ ground-state energy difference between perpendicular and parallel molecular
    orientation with respect to the magnetic field direction as a function of field strength. The horizontal black line
    marks zero energy difference.\label{orientation}}
\end{center}
\end{figure}

Magnetic optical activity calculations are performed for twenty magnetic-field strengths in the range $0\,\text{T}$ to $55\,000\,\text{T}$ by extracting
the imaginary part of the polarizability as described in Sec.~\ref{subsec:extract}.
The field strength of the time-dependent electric field is chosen to be $\mathcal{E} = 0.001\,\text{a.u.}$, which is small enough to
warrant the dipole expansion in Eq.~\eqref{eq:dipole_expansion}.
We consider two cases: The orientation of the molecule is either taken to be independent of the direction of the magnetic field---i.e., the molecules of the sample are assumed to be randomly oriented---and the MOR is computed according to Eq.~\eqref{free_mor},
or the molecule is taken to be \emph{fixed} at the energetically most favorable orientation with respect to the magnetic-field direction and
Eq.~\eqref{fixed_mor} is used.
To determine the most favorable orientation, the Hartree--Fock/aug-cc-pVDZ ground-state bond lengths were computed at both the perpendicular and parallel orientation for each molecule. The ground-state energy difference as a function of the magnetic-field strength is displayed in Figure~\ref{orientation}.
For H$_2$ and HF, the parallel orientation with respect to the magnetic field direction is the most energetically favorable, while the perpendicular
orientation is found to be most favorable for CO.
All three molecules are closed-shell diamagnetic molecules, and the difference in orientation stems
from differences in the terms quadratic in the magnetic field---the diamagnetic terms---contained in Eq.~\eqref{eq:magnetic_hamiltonian}.
The optimized bond lengths for the three molecules can be found in the Supplementary Information. 

When evaluating Eq.~\eqref{free_mor} and Eq.~\eqref{fixed_mor}, the ideal gas approximation is employed at standard temperature ($0\,^\circ\text{C}$) and pressure ($1\,\text{atm}$). The number density thus is
$\mathcal{N} = 2.68678 \times 10^{25}\,\text{m}^{-3} = 3.98134 \times 10^{-6}\,\text{a}_0^{-3}$, and
\begin{equation}
    C = 2.53048 \times 10^{27}\,\frac{\text{J}\,\text{s}}{\text{C}^2\,\text{m}^3}
    = 9.12748 \times 10^{-8} \frac{\hbar}{\text{e}^2\,\text{a}_0^3}.
\end{equation}
We compute the MOR for the H$_2$ molecule at $\omega = 0.08284\,\text{a.u.}$ ($\lambda=550\,\text{nm}$), while
$\omega = 0.11391\,\text{a.u.}$ ($\lambda= 400\,\text{nm}$) is used for the CO and HF molecules in accordance with the
experimental work of \citeauthor{ingersoll_lebenberg}.~\cite{ingersoll_lebenberg} 

For the randomly oriented case, simulations are carried with experimental field-free bond lengths: $1.4\,\text{a}_0$ for H$_2$,
$2.132\,\text{a}_0$ for CO, and $1.7328\,\text{a}_0$ for HF. For the fixed orientation case,
the bond lengths of H$_2$, CO, and HF are optimized for each magnetic-field strength at the cTPSS/aug-cc-pVDZ level of theory
with the magnetic field vector parallel (H$_2$ and HF) or perpendicular (CO) to the bond axis.

We compute the MOR from TDHF, TDCC2, TDCCSD, TDOMP2, and TDNOCCD simulations with the HyQD software.~\cite{HyQD}
Apart from using a stricter residual norm convergence criteria of $10^{-12}$ for the implicit equations of the Gauss-Legendre integrator, 
the same convergence thresholds are applied for the MOR calculations as those specified in Sec.~\ref{sec:absspec} 
The TDCDFT simulations are performed with the current--dependent exchange--correlation functionals cTPSS, cTPSSh, and cTPSSrsh
as described in Ref.~\onlinecite{wibowo_modeling_2021}. The density is propagated using the Magnus 2 propagator~\cite{blanes2016concise}
with a modest time step of $0.1\,\text{a.u.}$, which has been shown to yield a good balance between accuracy and
efficiency for computing absorption spectra in the presence of a magnetic field.~\cite{wibowo_modeling_2021}
The TDCDFT simulations are performed with the \textsc{QUEST} code.~\cite{QUEST} For all calculations, the aug-cc-pVDZ basis set is employed,
and it should be noted that the basis set limit has not been reached.~\cite{coriani_gauge-origin_2000} Moreover, the CC expansion lacks triple excitations, which are important for high accuracy,~\cite{coriani_triple_2000} and rovibrational effects are not taken into account.
Consequently, the simulations presented here cannot be expected to reproduce or predict experimental results with very high accuracy.

\begin{figure*}[h]
\begin{center}
\includegraphics[scale=0.73]{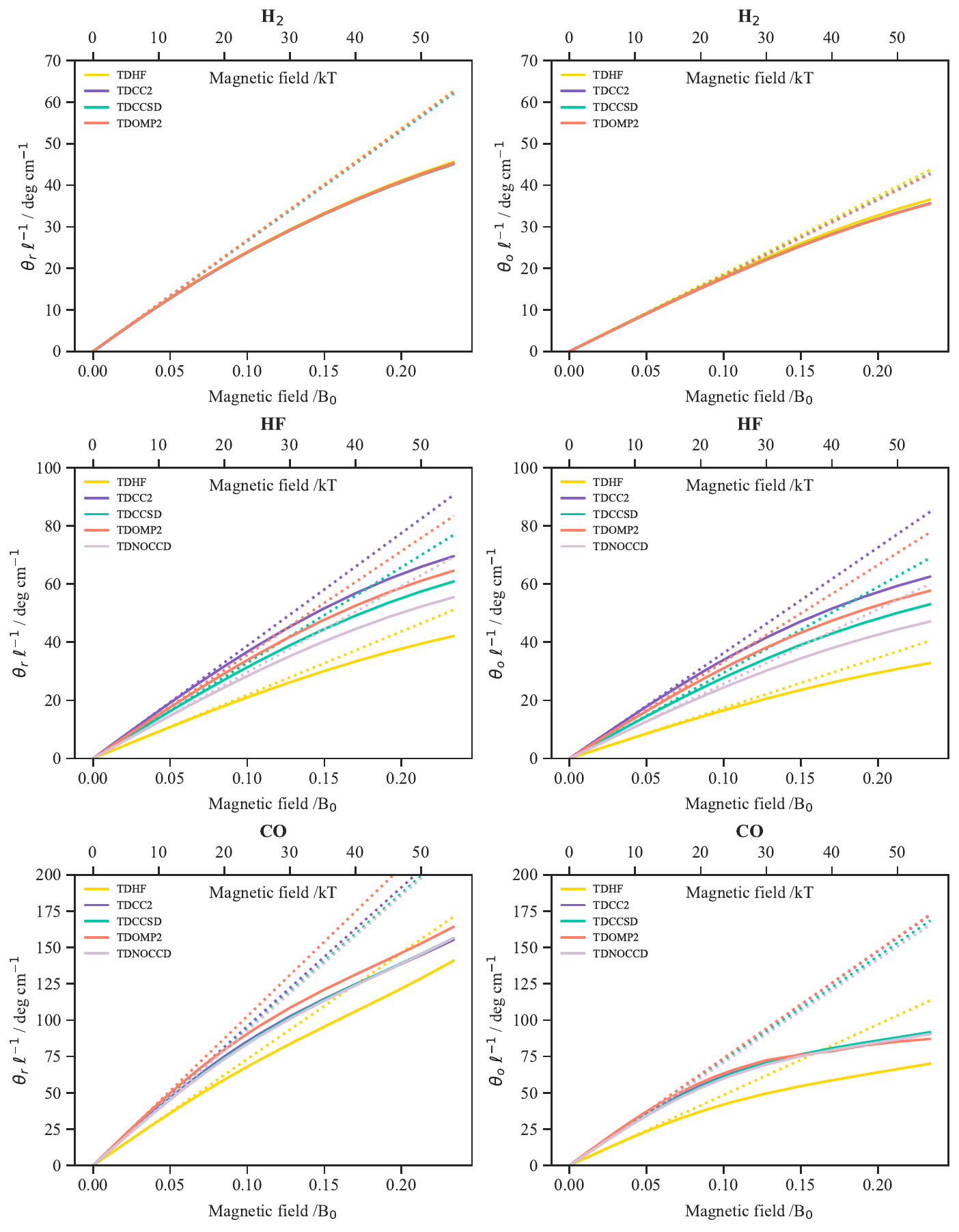} 
    \caption{The magnetic optical rotation of H$_2$, HF, and CO as a function of the magnetic-field strength computed with wave function methods. The panels on the left display the MOR for the randomly oriented molecules, while the panels on the right display the MOR for the 
    fixed orientation relative to the magnetic field vector (parallel for H$_2$ and HF systems, perpendicular for CO). The dashed lines are
    obtained from the Verdet constants in Table \ref{verdet_values}.\label{wave_fig}}
\end{center}
\end{figure*}

\subsubsection{Verdet constants}

We determine Verdet constants by fitting the MOR data for randomly oriented samples to a fifth-order polynomial in the magnetic-field strength,
identifying $V(\omega)$ as the coefficient of the linear term.
The results are listed in Table \ref{verdet_values} along with the experimental results of
\citeauthor{ingersoll_lebenberg}.~\cite{ingersoll_lebenberg}
\begin{table}[h]
\begin{center}
    \caption{\label{verdet_values}The Verdet constant (in $10^{-7}\,\text{a.u.}$) extracted from simulations.
    Experimental values are taken from Ref.~\onlinecite{ingersoll_lebenberg}.}
\begin{threeparttable}
\begin{tabular}{ l  r  r  r }
\hline
\hline
      & H$_2$ & HF & CO \\
$\omega/\text{a.u.}$ & 0.08284 & 0.11391 & 0.11391\\
\hline
TDHF     & 0.248 & 0.202 & 0.687 \\
TDCC2    & 0.246 & 0.358 & 0.906 \\
TDCCSD   & 0.245 & 0.304 & 0.884 \\
TDOMP2   & 0.247 & 0.329 & 0.972 \\
TDNOCCD  & -     & 0.274 & 0.876 \\
cTPSS    & 0.303 & 0.416 & 0.987 \\
cTPSSh   & 0.235 & 0.288 & 0.761 \\
cTPSSrsh & 0.160 & 0.202 & 0.512 \\
Exp.     & 0.251 & -     & 0.895 \\
\hline
\hline
\end{tabular}
\end{threeparttable}
\end{center}
\end{table}

While electron correlation effects are important for HF and CO, their impact is less pronounced in the case of H$_2$. The wave function methods that take electron correlation into account yield Verdet constant which agree with experimental values (for H$_2$ and CO) to within $2$--$10\%$, which is reasonable considering
the lack of higher-order correlation effects~\cite{coriani_triple_2000}
and vibrational contributions,~\cite{mort_vibrational_2007} in addition to the relatively small basis set. With errors ranging from $6\%$ to $38\%$, the TDCDFT model with the cTPSSrsh functional exhibit comparatively large errors, especially considering that the TDHF results remain below $24\%$.

Computations conducted using quadratic response theory~\cite{parkinson_calculation_1993, coriani_gauge-origin_2000} should yield identical Verdet constants, although slight variations may arise, for example, due to numerical errors from the finite-difference calculations involved in
the linear-response-function extraction.~\cite{ding_efficient_2013,ofstad_adiabatic_2023}
\citeauthor{parkinson_calculation_1993}~\cite{parkinson_calculation_1993} computed Verdet constants for the three molecules at the Hartree--Fock level (i.e., the random phase approximation). They used a somewhat larger basis set than aug-cc-pVDZ, albeit \emph{without} the LAO phase factors, and obtained Verdet constants
of $0.232 \times 10^{-7}\,\text{a.u.}$ for H$_2$,
$0.219 \times 10^{-7}\,\text{a.u.}$ for HF, and
$0.702 \times 10^{-7}\,\text{a.u.}$ for CO, in fair agreement with our TDHF results in Table \ref{verdet_values}. \citeauthor{coriani_gauge-origin_2000}~\cite{coriani_gauge-origin_2000} computed the Verdet constant for HF at the CCSD level with the same basis set to be $0.3038 \times 10^{-7}\,\text{a.u.}$, which is within $0.03\%$ of our result ($0.3037 \times 10^{-7}\,\text{a.u.}$). This confirms that our MOR procedure indeed
reproduces quadratic response theory at low magnetic-field strengths.

\subsubsection{TDCC results at finite magnetic field}

Figure \ref{wave_fig} shows the MOR obtained from wave function-based simulations of H$_2$, HF, and CO
as a function of magnetic-field strength, with the dashed lines
representing the values predicted from Eq.~\eqref{eq:verdet} using the Verdet constants reported in Table \ref{verdet_values}.
Results for randomly oriented molecules are shown in the left-hand panels, while those for oriented molecules are
shown in the right-hand panels. There are significant deviations from linearity for both oriented and randomly oriented samples,
as we shall discuss in more detail below.

The two uppermost panels of Fig.~\ref{wave_fig} show the MOR of H$_2$ obtained
from TDHF, TDCC2, TDCCSD, and TDOMP2 simulations.
Since the H$_2$ molecule is a two-electron system, the TDCCSD and TDNOCCD methods are equivalent.
As also observed for the Verdet constants above, electron correlation effects are not highly important for the MOR of H$_2$ at the
range of magnetic-field strengths considered here. The effect of orientation---which includes varying bond lengths---is
significant, with $\theta_r$ roughly $28\%$ greater than $\theta_o$.

For both HF and CO, however, correlation effects are crucial, confirming previous observations based on quadratic response theory by
\citeauthor{parkinson_calculation_1993}~\cite{parkinson_calculation_1993} and
\citeauthor{coriani_triple_2000}~\cite{coriani_triple_2000}
While the TDCCSD and TDNOCCD results agree for CO, the optimization of the time-dependent orbitals in the TDNOCCD method give rise to a significant change
compared with the static orbitals of TDCCSD theory for the HF molecule. This observation also applies to the second-order approximations
TDCC2 and TDOMP2, where we also note that TDOMP2 results are closer to the TDCCSD ones than those obtained from TDCC2 theory for HF. However, this is not at all the case for CO.
The latter is in agreement with magnetic field-free polarizability and hyperpolarizability results reported by
\citeauthor{omp2}~\cite{omp2}
It is, however, not possible to conclude which of the TDNOCCD and TDCCSD methods is superior based on the present results,
as it requires a more careful study of convergence with respect to higher-order excitations beyond doubles.
We will, therefore, compare TDCDFT results with the MOR obtained from both the TDCCSD and TDNOCCD methods.

It is noteworthy that the MOR is almost identical
for oriented and randomly oriented HF, whereas a significant orientation effect is observed for the CO molecule.
While one might speculate that the ``oriented'' contribution, Eq.~\eqref{fixed_mor}, dominates the averaging in Eq.~\eqref{free_mor} for HF,
it turns out to be caused by the change in bond length obtained in the magnetic-field dependent geometry optimization.

\begin{figure*}[h]
\begin{center}
\includegraphics[scale=0.73]{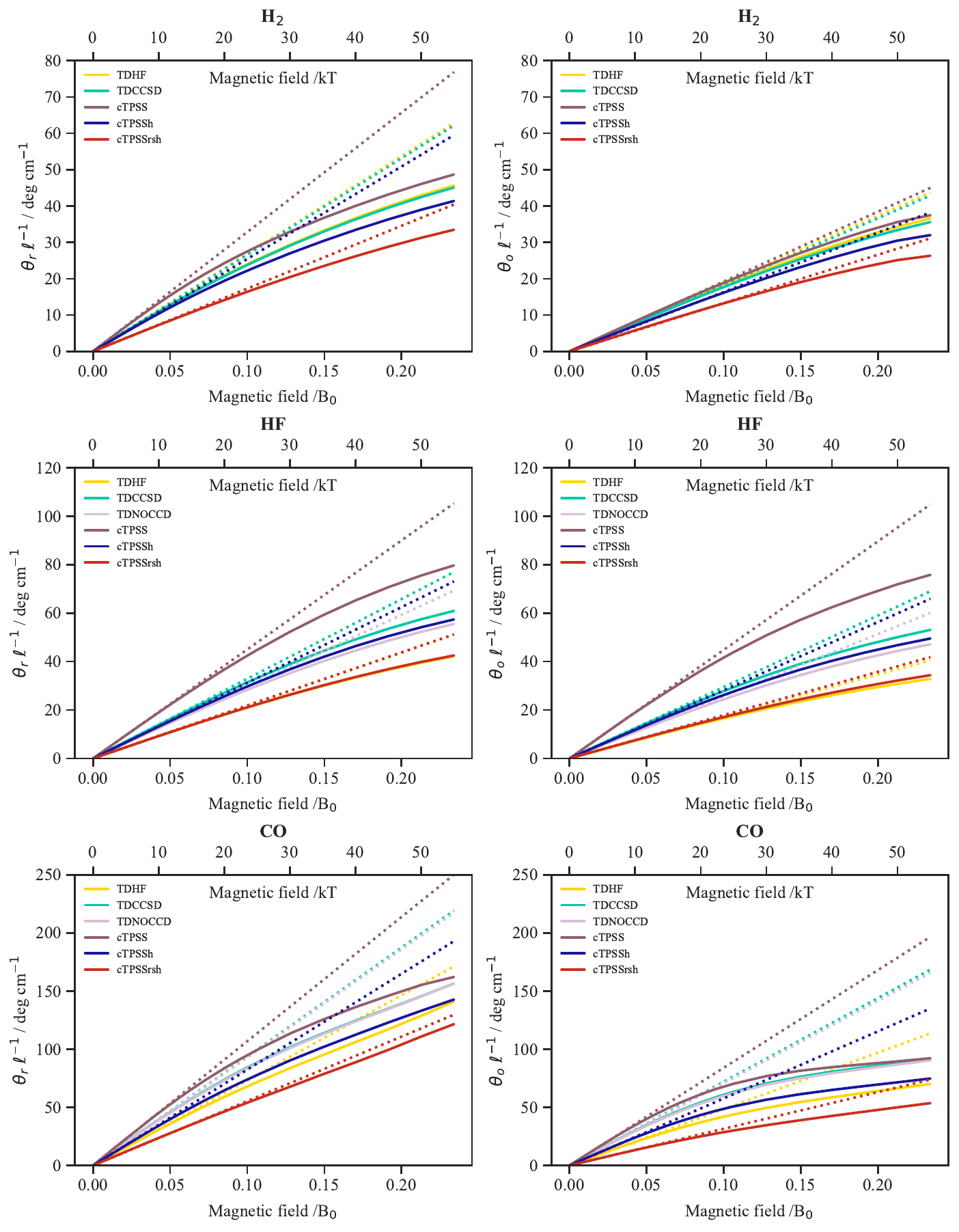} 
    \caption{The magnetic optical rotation of H$_2$, HF, and CO as a function of the magnetic-field strength computed with TDCDFT and the TDHF, TDCCSD, and TDNOCCD methods. 
    The panels on the left display the MOR for the randomly oriented molecules, while the panels on the right display the MOR for the 
    fixed orientation relative to the magnetic field vector (parallel for H$_2$ and HF systems, perpendicular for CO). The dashed lines are
    obtained from the Verdet constants in Table \ref{verdet_values}.\label{dft_fig}}
\end{center}
\end{figure*}

\subsubsection{TDCDFT results at finite magnetic field}

Figure \ref{dft_fig} shows the MOR obtained from TDCDFT simulations of H$_2$, HF, and CO
as a function of magnetic-field strength with the dashed lines
representing the predicted values using Eq.~\eqref{eq:verdet} with the Verdet constants reported in Table \ref{verdet_values}.
Results for randomly oriented and oriented molecules are shown in the left- and right-hand panels, respectively, and
the TDHF, TDCCSD, and TDNOCCD results are reproduced here for comparison.
Also for the TDCDFT methods, we observe significant deviations from linearity as the magnetic-field strength increases.

The cTPSSh and cTPSSrsh functionals have previously been applied to the calculation of isotropic NMR shielding
constants,~\cite{irons_analyzing_2020} where no significant improvements were found compared with the cTPSS functional.
The cTPSSrsh functional has, however, been found to better describe excited states compared to cTPSS and cTPSSh in the magnetic field-free
case.~\cite{wibowo_modeling_2021} We observe that the cTPSSh functional generally produces MOR values that are somewhat closer to TDCCSD/TDNOOCD results
than the cTPSS functional, Moreover, the cTPSSh functional outperforms the TDHF method for HF, and CO, but not for H$_2$.
The MOR values computed using the cTPSSh functional fall in between the cTPSS and cTPSSrsh results,
which are over- and under-estimates of the MOR, respectively.
The poor performance of the cTPSSrsh functional is somewhat surprising considering that range separation is known 
to improve the description of nonlinear properties.\cite{garza_assessment_2013, choluj_benchmarking_2018}
The cTPSSrsh functional  incorporates a large fraction of the current-independent PBE functional, which might contribute to the quite poor performance
in finite magnetic fields. When the magnetic field is set to zero, the cTPSSrsh functional yields ground state energies and dipole moments comparable to cTPSSh, both of which outperform the cTPSS functional.

\subsubsection{Deviation from Verdet's linear law}

\begin{table}[t]
\caption{\label{diverge_values}The magnetic-field strength (in kT) at which the nonperturbative MOR deviates from Verdet's law
    by a chosen percentage.}
\begin{center}
\begin{tabular}{llrrrr}
\hline
\hline
      &          & \multicolumn{4}{c}{Orientation} \\
      &          & \multicolumn{2}{c}{Random} & \multicolumn{2}{c}{Fixed} \\
      &          &  3\%  &\quad  5\%  &  \quad \quad 3\%  & \quad 5\%   \\
\hline
H$_2$ & TDHF     & 9.03 & 13.1 & 22.1 & 28.5 \\
      & TDCC2    & 9.14 & 13.2 & 21.9 & 28.2 \\
      & TDCCSD   & 9.25 & 13.3 & 21.6 & 27.8 \\
      & TDOMP2   & 9.03 & 13.0 & 21.9 & 28.2 \\
      & cTPSS    & 6.39 & 9.69 & 27.1 & 32.3 \\
      & cTPSSh   & 7.76 & 11.5 & 29.1 & 34.0 \\
      & cTPSSrsh & 16.2 & 22.6 & 31.2 & 36.2 \\
HF    & TDHF     & 20.3 & 26.4 & 18.7 & 24.4 \\
      & TDCC2    & 17.1 & 22.1 & 15.4 & 20.0 \\
      & TDCCSD   & 18.7 & 24.0 & 16.2 & 21.3 \\
      & TDOMP2   & 17.5 & 22.6 & 15.1 & 19.5 \\
      & TDNOCCD  &\quad 19.3 & 24.7 & 18.3 & 23.2 \\
      & cTPSS    & 16.9 & 21.4 & 15.2 & 19.3 \\
      & cTPSSh   & 18.1 & 23.3 & 13.3 & 18.6 \\
      & cTPSSrsh & 21.2 & 27.4 & 18.2 & 24.4 \\
CO & TDHF & 14.9 & 19.0  & 11.6 & 14.2 \\
      & TDCC2 & 12.4 & 15.4 & 10.7 & 12.9 \\
      & TDCCSD & 12.8 & 16.0  & 11.0 & 13.3 \\
      & TDOMP2 & 11.5 & 14.5  & 10.2 & 11.4 \\
      & TDNOCCD & 12.7 & 15.9 &  11.0 & 13.3 \\
      & cTPSS & 13.7 & 16.2 &  9.86 & 11.9 \\
      & cTPSSh & 11.0 & 14.6  & 11.1 & 13.4 \\
      & cTPSSrsh & 24.9 & 33.3  & 13.9 & 17.0 \\
\hline
\hline
\end{tabular}
\end{center}
\end{table}

As noted above,
it is evident from Figs.~\ref{wave_fig} and \ref{dft_fig} that Verdet's linear relation between the MOR and the magnetic-field strength
only holds for sufficiently small magnetic-field strengths. While this is not surprising, considering that the observations of Verdet
can be accounted for by first-order perturbation theory, it is not straightforward to uniquely define a field strength at which
linearity is broken. In this work, we will define the ``breaking point'' as the magnetic-field strength at which
the fifth-order polynomial function fitted to the nonperturbative MOR deviates from the value predicted from the Verdet constants, defined as the linear coefficient of this polynomial function, by a given percentage. 

The magnetic-field strengths at which $3\%$ and $5\%$ deviations are observed can be found in Table \ref{diverge_values}.
With random orientation, the breaking point roughly lies somewhere between $10$ and $20\,\text{kT}$ ($0.04$--$0.09\,\text{B}_0$),
$2$--$3$ orders of magnitude
above the strongest sustainable static magnetic-field strength produced on Earth with a high-temperature superconducting magnet, namely
$45.5\,\text{T}$ ($2 \times 10^{-4}\,\text{B}_0$).~\cite{hahn_455-tesla_2019}
At $45.5\,\text{T}$, the TDCCSD and TDNOCCD results deviate from linearity by about $0.9\%$ for H$_2$ and CO and by about $0.09\%$ for HF
(with random orientation). Not surprisingly, our results thus indicate that Verdet's law can be safely applied for the range of
magnetic-field strengths that can be produced in experimental setups.

\section{Concluding remarks}\label{sec:conc}

We have in this work presented the first implementation of TDCC theory
for the study of laser-driven multi-electron dynamics in finite magnetic fields. The implementation is validated
by comparing linear absorption spectra obtained from simulations with those obtained from time-independent EOM-EE-CC
theory. The implementation supports both static and dynamic reference determinants with the orbitals expanded
in London atomic orbitals to ensure magnetic gauge-origin invariance.

The new implementation is applied to the calculation of magnetic optical rotation at finite magnetic-field strengths, demonstrating 
that Verdet's linear relationship is valid up to field strengths of roughly $10$--$20\,\text{kT}$ ($0.04$--$0.09\,\text{B}_0$).
A deviation from linearity below $1\%$ is found at $45.5\,\text{T}$, which is the strongest sustainable static magnetic field
ever produced by a superconducting magnet on Earth.
We have also compared the TDCC magnetic optical rotations with those obtained from TDCDFT simulations using the current-dependent
cTPSS, cTPSSh, and cTPSSrsh density-functional approximations. With TDCCSD and TDNOCCD results as benchmark, we find that the
best-performing functional is cTPSSh. The cTPSS functional tends to overestimate the rotation while the cTPSSrsh functional tends to
underestimate it.

\section*{Supplementary material}\label{sec:supmat}

The supplementary material contains cTPSS/aug-cc-pVDZ equilibrium bond lengths of H$_2$, HF, and CO
placed in a magnetic field parallel or perpendicular to the bond axis with field strengths ranging from $1\,\text{T}$ to $55\,\text{kT}$.
The overlayed absorption spectra generated from TDCC2 simulations and from time-independent EOM-EE-CC2 calculations can also be found
in the supplementary material.

\section*{Acknowledgment}\label{sec:ack}

This work was supported by the Research Council of Norway through its Centres of Excellence scheme, project number 262695. The calculations were performed on resources provided by Sigma2---the National Infrastructure for High Performance Computing and Data Storage in Norway, Grant No.~NN4654K. SK and TBP acknowledge the support of the Centre for Advanced Study in Oslo, Norway, which funded and hosted the CAS research project \emph{Attosecond Quantum Dynamics Beyond the Born-Oppenheimer Approximation} during the academic year 2021-2022.

\section*{Data availability statement}

The data that support the findings of this study are available from the corresponding author upon reasonable request.

\bibliography{Manuscript}

\end{document}


\title{Supplementary material for ``Magnetic Optical Rotation from Real-Time Simulations in Finite Magnetic Fields''}
\author{Benedicte Sverdrup Ofstad}
\affiliation{\footnotesize Hylleraas Centre for Quantum Molecular Sciences, Department of Chemistry, 
             University of Oslo, 0315 Oslo, Norway}
\author{Meilani Wibowo}
\affiliation{\footnotesize School of Chemistry, University of Nottingham, University Park, Nottingham NG7 2RD, United Kingdom}  
\author{H{\aa}kon Emil Kristiansen}
\affiliation{\footnotesize Hylleraas Centre for Quantum Molecular Sciences, Department of Chemistry, 
             University of Oslo, 0315 Oslo, Norway}
\author{Einar Aurbakken}
\affiliation{\footnotesize Hylleraas Centre for Quantum Molecular Sciences, Department of Chemistry, 
             University of Oslo, 0315 Oslo, Norway}
\author{Marios Petros Kitsaras}
\affiliation{\footnotesize Department Chemie, Johannes Gutenberg-Unversität Mainz, Duesbergweg 10-14, D-55128 Mainz,
Germany}  
\author{{\O}yvind Sigmundson Sch{\o}yen}
\affiliation{\footnotesize Department of Physics, University of Oslo, Norway}      
\author{Eirill Hauge}
\affiliation{\footnotesize Department of Numerical Analysis and Scientific Computing, University of Oslo, Norway}     
\author{Tom J. P. Irons}
\affiliation{\footnotesize School of Chemistry, University of Nottingham, University Park, Nottingham NG7 2RD, United Kingdom}      
\author{Simen Kvaal}
\affiliation{\footnotesize Hylleraas Centre for Quantum Molecular Sciences, Department of Chemistry, 
             University of Oslo, 0315 Oslo, Norway}        
\author{Stella Stopkowicz}
\affiliation{\footnotesize Department Chemie, Johannes Gutenberg-Unversität Mainz, Duesbergweg 10-14, D-55128 Mainz,
Germany}
\author{Andrew Teale}
\affiliation{\footnotesize Hylleraas Centre for Quantum Molecular Sciences, Department of Chemistry, 
             University of Oslo, 0315 Oslo, Norway}
\affiliation{\footnotesize School of Chemistry, University of Nottingham, University Park, Nottingham NG7 2RD, United Kingdom}    
\author{Thomas Bondo Pedersen}
\affiliation{\footnotesize Hylleraas Centre for Quantum Molecular Sciences, Department of Chemistry, 
             University of Oslo, 0315 Oslo, Norway}

\maketitle

\section{Absorption plots}

\begin{figure}[H]
\begin{center}
\begin{subfigure}{0.44\textwidth}
\includegraphics[width=\textwidth]{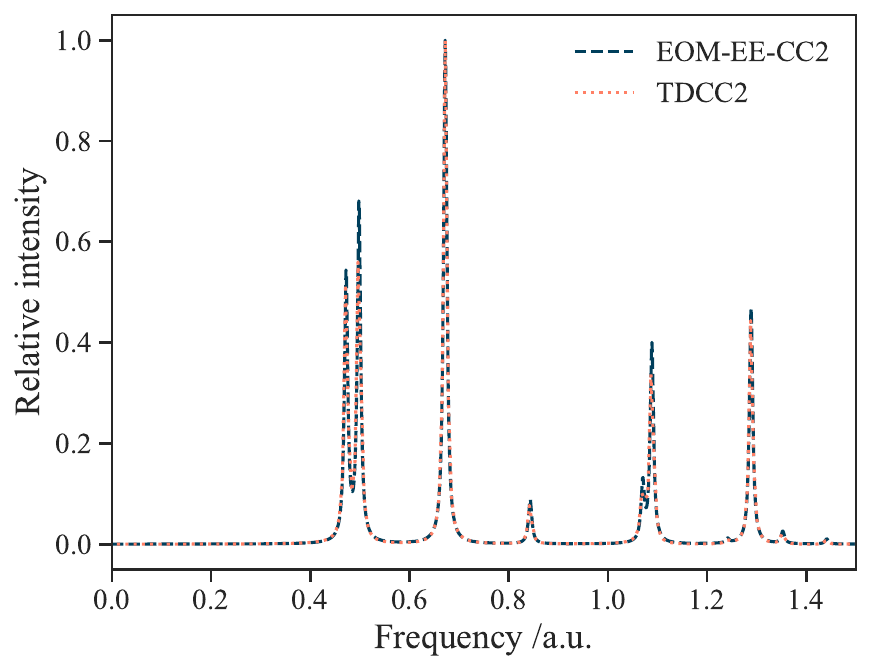} 
\caption{\centering H$_2$}
\label{abs_h2}
\end{subfigure}
\begin{subfigure}{0.44\textwidth}
\includegraphics[width=\textwidth]{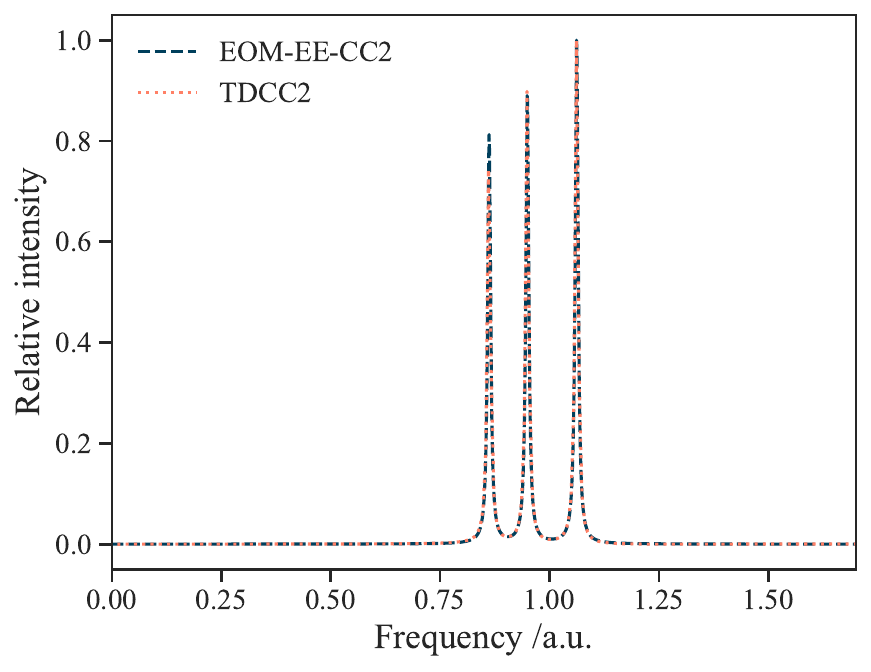} \\
\caption{\centering He}
\end{subfigure}
\begin{subfigure}{0.44\textwidth}
\includegraphics[width=\textwidth]{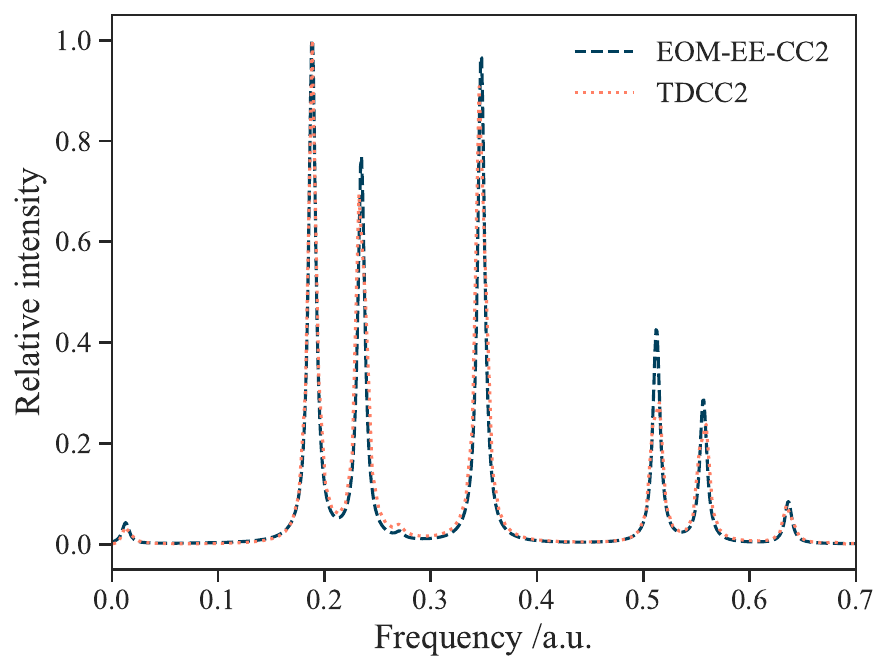} 
\caption{\centering Be}
\label{sec_hyp_thg}
\end{subfigure}
\begin{subfigure}{0.44\textwidth}
\includegraphics[width=\textwidth]{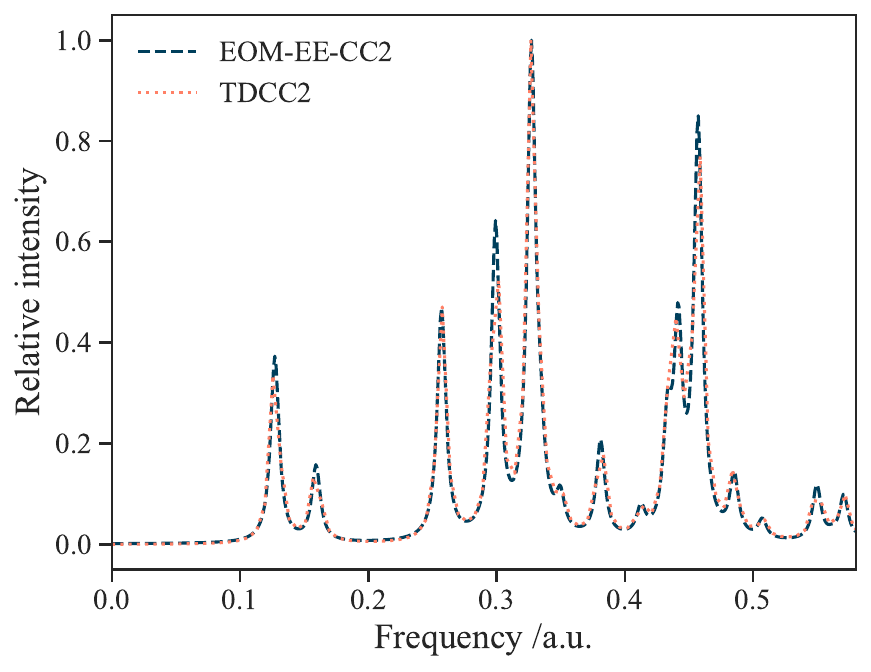} 
\caption{\centering LiH}
\end{subfigure}
    \caption{Overlay of absorption spectra generated by TDCC2 and EOM-EE-CC2 at the magnetic field strength $0.2\,\text{B}_0$ (along the
    internuclear axis for the diatomic molecules) with the aug-cc-pVTZ basis set. \label{absorption_plots_cc2}}
\end{center}
\end{figure}

The comparison of absorption spectra generated with TDCC2 and EOM-EE-CC2 are found in Fig.~\ref{absorption_plots_cc2}.

\newpage

\section{Bond Lengths}

The optimized bond lengths and ground state energies of H$_2$ and HF and CO used in the magnetic optical rotation calculations for fixed orientation case are found in Table \ref{bond_lengths}. The geometry optimization is evaluated at the CDFT level of theory using the cTPSS functional and the aug-cc-pVDZ basis set as implemented in \textsc{QUEST}. The magnetic field is parallel to the bond axis for H$_2$ and HF, and 
perpendicular to the bond axis for CO.\\

{
\centering
\bottomcaption{The optimized bond lengths of H$_2$, HF, and CO in magnetic fields.  \label{bond_lengths}}
\tablefirsthead{
\hline
\hline \\
& \quad Magnetic field/ T&  \quad \quad \quad   Bond length /Å&   \quad \quad \quad SCF Energy / $E_h$\\
}
\tablehead{
&\quad Magnetic field/ T&  \quad \quad \quad   Bond length /Å&   \quad \quad \quad SCF Energy / $E_h$\\
 \ &  \\}
\tablelasttail{  \\ \hline  \hline}
\begin{supertabular}{r   r   r   r  }  
H$_2$ \\
& 1       & 0.757591664716 & -1.128763813026\\ 
& 1000 & 0.757587852164 & -1.128756840035\\ 
& 3000 & 0.757557362818 & -1.128701061953\\ 
& 5000 & 0.757496443178 & -1.128589537228\\ 
& 7000  & 0.757405210630 & -1.128422328514\\ 
& 9000   & 0.757283839520 & -1.128199529246\\ 
& 11000 & 0.757132559100 & -1.127921262997\\ 
& 13000 & 0.756951650934 & -1.127587682646\\ 
& 15000 & 0.756741445760 & -1.127198969369\\ 
& 17000 & 0.756502319988 & -1.126755331481\\ 
& 20000 & 0.756090330516 & -1.125987409979\\ 
& 22500 & 0.755699149368 & -1.125254120646\\ 
& 25000 & 0.755265516964 & -1.124436575385\\ 
& 30000 & 0.754275178002 & -1.122551266357\\ 
& 35000 & 0.753128327708 & -1.120337003495\\ 
& 40000 & 0.751834478186 & -1.117799883198\\ 
& 45000 & 0.750403368646 & -1.114946447882\\ 
& 55000 & 0.748844772680 & -1.111783575794\\
HF \\
&1 & 0.933590956442 & -100.032489121335\\ 
&1000 & 0.933589135020 & -100.032469938912\\ 
&3000 & 0.933574566434 & -100.032316484574\\ 
&5000 & 0.933545444164 & -100.032009603634\\ 
&7000 & 0.933501797946 & -100.031549351471\\ 
&9000 & 0.933443672214 & -100.030935810935\\ 
&11000 & 0.933371125897 & -100.030169092079\\ 
&13000 & 0.933284232154 & -100.029249331811\\ 
&15000 & 0.933183078039 & -100.028176693459\\ 
&17000 & 0.933067764110 & -100.026951366266\\ 
&20000 & 0.932868495579 & -100.02482756003\\ 
&22500 & 0.932678611471 & -100.022796254145\\ 
&25000 & 0.932467367754 & -100.020527826219\\ 
&30000 & 0.931982068349 & -100.015282046434\\ 
&35000 & 0.931415364343 & -100.00909561215\\ 
&40000 & 0.930770345220 & -100.001974659863\\ 
&45000 & 0.930050347468 & -99.993925973387\\ 
&55000 & 0.929258877253 & -99.984956883481 \\
CO \\
& 1 & -1.143707581573 & -112.752788860962\\ 
& 1000 & -1.143705667323 & -112.752754533032\\ 
& 3000 & -1.143690355574 & -112.752479925779\\ 
& 5000 & -1.143659744145 & -112.751930798993\\ 
& 7000 & -1.143613857074 & -112.751107327681\\ 
& 9000 & -1.143552730182 & -112.750009773271\\ 
& 11000 & -1.143476410785 & -112.748638482324\\ 
& 13000 & -1.143384957333 & -112.746993884863\\ 
& 15000 & -1.143278438979 & -112.745076492332\\ 
& 17000 & -1.143156935045 & -112.742886895217\\ 
& 20000 & -1.142946778279 & -112.739093594855\\ 
& 22500 & -1.142746288046 & -112.735467653153\\ 
& 25000 & -1.142522974387 & -112.731420880667\\ 
& 30000 & -1.142008832556 & -112.722072279283\\ 
& 35000 & -1.141406410228 & -112.711064060774\\ 
& 40000 & -1.140734386537 & -112.698411879142\\ 
& 45000 & -1.139959487538 & -112.684141290936\\ 
& 55000 & -1.139103510635 & -112.668270458767\\ 
\end{supertabular}
}

\newpage

\newgeometry{top=25mm, bottom=20mm, hmargin=5mm} 
\section{Magnetic optical rotations}

\begin{table}[h]
\begin{center}
\caption{The magnetic optical rotation computed for H$_2$. Basis set: aug-cc-pVDZ basis set. Both the magnetic fields and optical rotations are reported in atomic units.}
\begin{tabular}{ l r r r r r r r r r }
\hline
\hline
\\
Magnetic field & TDHF &\quad  \quad TDCC2 & \quad  \quad TDCCSD &  \quad \quad TDOMP2  &\quad  \quad  \quad  cTPSS  &\quad  \quad cTPSSh &  \quad  \quad TPSSrsh  \\
\vspace{-0.2cm}\\
 \multicolumn{8}{c}{Free orientation}\\
\vspace{-0.2cm}\\
4.2553191e-06 & 1.0452E-13  & 1.0383E-13  & 1.0342E-13  & 1.0404E-13  & 1.2727E-13  & 9.8700E-14  & 6.7606E-14  \\
0.0042553191 & 1.0449E-10  & 1.0380E-10  & 1.0339E-10  & 1.0401E-10  & 1.2721E-10  & 9.8671E-11  & 6.7663E-11  \\
0.0127659574 & 3.1281E-10  & 3.1076E-10  & 3.0953E-10  & 3.1136E-10  & 3.8037E-10  & 2.9534E-10  & 2.0280E-10  \\
0.0212765957 & 5.1927E-10  & 5.1587E-10  & 5.1386E-10  & 5.1676E-10  & 6.3001E-10  & 4.8998E-10  & 3.3737E-10  \\
0.029787234 & 7.2269E-10  & 7.1806E-10  & 7.1541E-10  & 7.1924E-10  & 8.7425E-10  & 6.8128E-10  & 4.7106E-10  \\
0.0382978723 & 9.2214E-10  & 9.1636E-10  & 9.1301E-10  & 9.1782E-10  & 1.1112E-09  & 8.6820E-10  & 6.0363E-10  \\
0.0468085106 & 1.1170E-09  & 1.1101E-09  & 1.1059E-09  & 1.1116E-09  & 1.3390E-09  & 1.0498E-09  & 7.3480E-10  \\
0.0553191489 & 1.3068E-09  & 1.2988E-09  & 1.2942E-09  & 1.3003E-09  & 1.5570E-09  & 1.2254E-09  & 8.6422E-10  \\
0.0638297872 & 1.4912E-09  & 1.4823E-09  & 1.4772E-09  & 1.4836E-09  & 1.7647E-09  & 1.3950E-09  & 9.9185E-10  \\
0.0723404255 & 1.6701E-09  & 1.6604E-09  & 1.6547E-09  & 1.6616E-09  & 1.9624E-09  & 1.5585E-09  & 1.1178E-09  \\
0.085106383 & 1.9284E-09  & 1.9177E-09  & 1.9113E-09  & 1.9184E-09  & 2.2410E-09  & 1.7926E-09  & 1.3034E-09  \\
0.0957446809 & 2.1350E-09  & 2.1234E-09  & 2.1163E-09  & 2.1235E-09  & 2.4580E-09  & 1.9783E-09  & 1.4548E-09  \\
0.1063829787 & 2.3340E-09  & 2.3214E-09  & 2.3138E-09  & 2.3210E-09  & 2.6628E-09  & 2.1561E-09  & 1.6029E-09  \\
0.1276595745 & 2.7109E-09  & 2.6963E-09  & 2.6875E-09  & 2.6945E-09  & 3.0402E-09  & 2.4909E-09  & 1.8888E-09  \\
0.1489361702 & 3.0616E-09  & 3.0446E-09  & 3.0343E-09  & 3.0413E-09  & 3.3822E-09  & 2.8005E-09  & 2.1614E-09  \\
0.170212766 & 3.3876E-09  & 3.3674E-09  & 3.3554E-09  & 3.3627E-09  & 3.6968E-09  & 3.0882E-09  & 2.4183E-09  \\
0.1914893617 & 3.6891E-09  & 3.6649E-09  & 3.6509E-09  & 3.6589E-09  & 3.9873E-09  & 3.3546E-09  & 2.6601E-09  \\
0.2127659574 & 3.9657E-09  & 3.9369E-09  & 3.9204E-09  & 3.9297E-09  & 4.2545E-09  & 3.5994E-09  & 2.8856E-09  \\
\vspace{-0.2cm}\\
 \multicolumn{8}{c}{Fixed orientation}\\
\vspace{-0.2cm}\\
4.2553191e-06 & 7.3369E-14  & 7.1861E-14  & 7.1935E-14  & 7.1740E-14  & 7.6763E-14  & 6.5726E-14  & 5.3655E-14  \\
0.0042553191 & 7.3383E-11  & 7.1850E-11  & 7.1947E-11  & 7.1728E-11  & 7.6774E-11  & 6.5715E-11  & 5.3647E-11  \\
0.0127659574 & 2.2020E-10  & 2.1542E-10  & 2.1588E-10  & 2.1501E-10  & 2.3045E-10  & 1.9695E-10  & 1.6081E-10  \\
0.0212765957 & 3.6659E-10  & 3.5874E-10  & 3.5940E-10  & 3.5807E-10  & 3.8389E-10  & 3.2786E-10  & 2.6782E-10  \\
0.029787234 & 5.1245E-10  & 5.0150E-10  & 5.0236E-10  & 5.0065E-10  & 5.3640E-10  & 4.5848E-10  & 3.7464E-10  \\
0.0382978723 & 6.5762E-10  & 6.4347E-10  & 6.4461E-10  & 6.4248E-10  & 6.8801E-10  & 5.8836E-10  & 4.8080E-10  \\
0.0468085106 & 8.0177E-10  & 7.8451E-10  & 7.8582E-10  & 7.8326E-10  & 8.3865E-10  & 7.1725E-10  & 5.8620E-10  \\
0.0553191489 & 9.4479E-10  & 9.2435E-10  & 9.2585E-10  & 9.2282E-10  & 9.8797E-10  & 8.4528E-10  & 6.9089E-10  \\
0.0638297872 & 1.0864E-09  & 1.0628E-09  & 1.0645E-09  & 1.0610E-09  & 1.1361E-09  & 9.7208E-10  & 7.9455E-10  \\
0.0723404255 & 1.2265E-09  & 1.1997E-09  & 1.2014E-09  & 1.1978E-09  & 1.2825E-09  & 1.0972E-09  & 8.9705E-10  \\
0.085106383 & 1.4333E-09  & 1.4018E-09  & 1.4037E-09  & 1.3995E-09  & 1.4980E-09  & 1.2815E-09  & 1.0485E-09  \\
0.0957446809 & 1.6021E-09  & 1.5667E-09  & 1.5685E-09  & 1.5641E-09  & 1.6737E-09  & 1.4321E-09  & 1.1723E-09  \\
0.1063829787 & 1.7675E-09  & 1.7280E-09  & 1.7298E-09  & 1.7253E-09  & 1.8461E-09  & 1.5797E-09  & 1.2935E-09  \\
0.1276595745 & 2.0863E-09  & 2.0390E-09  & 2.0402E-09  & 2.0358E-09  & 2.1781E-09  & 1.8635E-09  & 1.5278E-09  \\
0.1489361702 & 2.3875E-09  & 2.3322E-09  & 2.3327E-09  & 2.3284E-09  & 2.4901E-09  & 2.1308E-09  & 1.7499E-09  \\
0.170212766 & 2.6691E-09  & 2.6058E-09  & 2.6051E-09  & 2.6020E-09  & 2.7810E-09  & 2.3806E-09  & 1.9572E-09  \\
0.1914893617 & 2.9290E-09  & 2.8581E-09  & 2.8558E-09  & 2.8539E-09  & 3.0485E-09  & 2.6104E-09  & 2.1488E-09  \\
0.2127659574 & 3.1662E-09  & 3.0876E-09  & 3.0833E-09  & 3.0831E-09  & 3.2884E-09  & 2.8153E-09  & 2.3204E-09  \\
0.2340425532 & 3.3796E-09  & 3.2934E-09  & 3.2869E-09  & 3.2888E-09  & 3.4610E-09  & 2.9553E-09  & 2.4306E-09  \\
\hline
\end{tabular}    
\end{center}
\end{table}

\begin{table}[h]
\begin{center}
\caption{The magnetic optical rotation computed for HF. Basis set: aug-cc-pVDZ basis set. Both the magnetic fields and optical rotations are reported in atomic units.}
\begin{tabular}{ l r r r r r r r r r r }
\hline
\hline
\\
Magnetic field & TDHF &\quad  \quad TDCC2 & \quad  \quad TDCCSD &  \quad  \quad TDOMP2  & \quad  \quad NOCCD& \quad  \quad  \quad  cTPSS  &\quad  \quad cTPSSh &  \quad  \quad TPSSrsh  \\
\vspace{-0.2cm}\\
 \multicolumn{9}{c}{Free orientation}\\
\vspace{-0.2cm}\\
4.2553191e-06 & 8.5853E-14  & 1.5258E-13  & 1.2958E-13  & 1.3994E-13  & 1.1652E-13  & 1.7666E-13  & 1.2329E-13  & 8.5872E-14  \\
0.0042553191 & 8.5846E-11  & 1.5255E-10  & 1.2955E-10  & 1.3995E-10  & 1.1651E-10  & 1.7664E-10  & 1.2326E-10  & 8.5864E-11  \\
0.0127659574 & 2.5738E-10  & 4.5700E-10  & 3.8800E-10  & 4.1997E-10  & 3.4920E-10  & 5.2966E-10  & 3.6904E-10  & 2.5758E-10  \\
0.0212765957 & 4.2846E-10  & 7.5981E-10  & 6.4527E-10  & 6.9935E-10  & 5.8115E-10  & 8.8192E-10  & 6.1317E-10  & 4.2910E-10  \\
0.029787234 & 5.9880E-10  & 1.0606E-09  & 9.0141E-10  & 9.7658E-10  & 8.1211E-10  & 1.2328E-09  & 8.5560E-10  & 5.9995E-10  \\
0.0382978723 & 7.6808E-10  & 1.3593E-09  & 1.1559E-09  & 1.2512E-09  & 1.0416E-09  & 1.5810E-09  & 1.0963E-09  & 7.6968E-10  \\
0.0468085106 & 9.3599E-10  & 1.6548E-09  & 1.4079E-09  & 1.5231E-09  & 1.2688E-09  & 1.9243E-09  & 1.3349E-09  & 9.3805E-10  \\
0.0553191489 & 1.1023E-09  & 1.9462E-09  & 1.6568E-09  & 1.7917E-09  & 1.4935E-09  & 2.2613E-09  & 1.5706E-09  & 1.1049E-09  \\
0.0638297872 & 1.2668E-09  & 2.2330E-09  & 1.9021E-09  & 2.0562E-09  & 1.7152E-09  & 2.5922E-09  & 1.8030E-09  & 1.2701E-09  \\
0.0723404255 & 1.4292E-09  & 2.5146E-09  & 2.1439E-09  & 2.3162E-09  & 1.9339E-09  & 2.9169E-09  & 2.0315E-09  & 1.4335E-09  \\
0.085106383 & 1.6684E-09  & 2.9256E-09  & 2.4984E-09  & 2.6964E-09  & 2.2548E-09  & 3.3893E-09  & 2.3663E-09  & 1.6745E-09  \\
0.0957446809 & 1.8633E-09  & 3.2575E-09  & 2.7854E-09  & 3.0035E-09  & 2.5147E-09  & 3.7691E-09  & 2.6375E-09  & 1.8713E-09  \\
0.1063829787 & 2.0536E-09  & 3.5793E-09  & 3.0642E-09  & 3.3011E-09  & 2.7677E-09  & 4.1360E-09  & 2.9002E-09  & 2.0635E-09  \\
0.1276595745 & 2.4194E-09  & 4.1882E-09  & 3.5942E-09  & 3.8659E-09  & 3.2512E-09  & 4.8262E-09  & 3.3976E-09  & 2.4324E-09  \\
0.1489361702 & 2.7628E-09  & 4.7441E-09  & 4.0873E-09  & 4.3832E-09  & 3.7009E-09  & 5.4561E-09  & 3.8593E-09  & 2.7808E-09  \\
0.170212766 & 3.0831E-09  & 5.2481E-09  & 4.5369E-09  & 4.8540E-09  & 4.1134E-09  & 6.0244E-09  & 4.2822E-09  & 3.1059E-09  \\
0.1914893617 & 3.3783E-09  & 5.6971E-09  & 4.9442E-09  & 5.2748E-09  & 4.4900E-09  & 6.5285E-09  & 4.6622E-09  & 3.4057E-09  \\
0.2127659574 & 3.6482E-09  & 6.0899E-09  & 5.3066E-09  & 5.6452E-09  & 4.8260E-09  & 6.9722E-09  & 5.0011E-09  & 3.6789E-09  \\
0.2340425532 & 3.8932E-09  & 6.4280E-09  & 5.6247E-09  & 5.9654E-09  & 5.1249E-09  & 7.3589E-09  & 5.2991E-09  & 3.9268E-09  \\
\vspace{-0.2cm}\\
 \multicolumn{9}{c}{Fixed orientation}\\
\vspace{-0.2cm}\\
4.2553191e-06 & 6.7859E-14  & 1.4253E-13  & 1.1572E-13  & 1.3140E-13  & 1.0082E-13  & 1.7431E-13  & 1.1086E-13  & 7.0395E-14  \\
0.0042553191 & 6.7896E-11  & 1.4258E-10  & 1.1572E-10  & 1.3130E-10  & 1.0085E-10  & 1.7441E-10  & 1.1081E-10  & 7.0354E-11  \\
0.0127659574 & 2.0378E-10  & 4.2798E-10  & 3.4715E-10  & 3.9214E-10  & 3.0287E-10  & 5.2516E-10  & 3.3123E-10  & 2.1064E-10  \\
0.0212765957 & 3.3948E-10  & 7.1201E-10  & 5.7790E-10  & 6.5077E-10  & 5.0497E-10  & 8.7877E-10  & 5.4897E-10  & 3.4995E-10  \\
0.029787234 & 4.7422E-10  & 9.9294E-10  & 8.0695E-10  & 9.0854E-10  & 7.0569E-10  & 1.2316E-09  & 7.6418E-10  & 4.8822E-10  \\
0.0382978723 & 6.0778E-10  & 1.2717E-09  & 1.0335E-09  & 1.1636E-09  & 9.0385E-10  & 1.5790E-09  & 9.7765E-10  & 6.2560E-10  \\
0.0468085106 & 7.4029E-10  & 1.5471E-09  & 1.2570E-09  & 1.4145E-09  & 1.0997E-09  & 1.9177E-09  & 1.1891E-09  & 7.6208E-10  \\
0.0553191489 & 8.7143E-10  & 1.8162E-09  & 1.4777E-09  & 1.6615E-09  & 1.2941E-09  & 2.2461E-09  & 1.3974E-09  & 8.9740E-10  \\
0.0638297872 & 1.0009E-09  & 2.0793E-09  & 1.6949E-09  & 1.9040E-09  & 1.4859E-09  & 2.5662E-09  & 1.6019E-09  & 1.0312E-09  \\
0.0723404255 & 1.1285E-09  & 2.3376E-09  & 1.9077E-09  & 2.1407E-09  & 1.6731E-09  & 2.8814E-09  & 1.8023E-09  & 1.1632E-09  \\
0.085106383 & 1.3155E-09  & 2.7149E-09  & 2.2192E-09  & 2.4838E-09  & 1.9467E-09  & 3.3405E-09  & 2.0932E-09  & 1.3572E-09  \\
0.0957446809 & 1.4678E-09  & 3.0159E-09  & 2.4692E-09  & 2.7606E-09  & 2.1680E-09  & 3.7033E-09  & 2.3262E-09  & 1.5151E-09  \\
0.1063829787 & 1.6165E-09  & 3.3032E-09  & 2.7105E-09  & 3.0279E-09  & 2.3821E-09  & 4.0498E-09  & 2.5516E-09  & 1.6693E-09  \\
0.1276595745 & 1.9000E-09  & 3.8417E-09  & 3.1693E-09  & 3.5244E-09  & 2.7884E-09  & 4.6931E-09  & 2.9782E-09  & 1.9667E-09  \\
0.1489361702 & 2.1656E-09  & 4.3296E-09  & 3.5888E-09  & 3.9745E-09  & 3.1619E-09  & 5.2645E-09  & 3.3728E-09  & 2.2476E-09  \\
0.170212766 & 2.4113E-09  & 4.7609E-09  & 3.9713E-09  & 4.3791E-09  & 3.5067E-09  & 5.7658E-09  & 3.7230E-09  & 2.5069E-09  \\
0.1914893617 & 2.6378E-09  & 5.1450E-09  & 4.3167E-09  & 4.7356E-09  & 3.8165E-09  & 6.2176E-09  & 4.0359E-09  & 2.7482E-09  \\
0.2127659574 & 2.8453E-09  & 5.4836E-09  & 4.6249E-09  & 5.0528E-09  & 4.0995E-09  & 6.6401E-09  & 4.3192E-09  & 2.9711E-09  \\
0.2340425532 & 3.0336E-09  & 5.7775E-09  & 4.9024E-09  & 5.3321E-09  & 4.3510E-09  & 7.0001E-09  & 4.5730E-09  & 3.1746E-09  \\
\hline
\end{tabular}    
\end{center}
\end{table}

\begin{table}[h]
\begin{center}
\caption{The magnetic optical rotation computed for CO. Basis set: aug-cc-pVDZ basis set. Both the magnetic fields and optical rotations are reported in atomic units.}
\begin{tabular}{ l r r r r r r r r r r }
\hline
\hline
\\
Magnetic field & TDHF &\quad  \quad TDCC2 & \quad  \quad TDCCSD &  \quad  \quad TDOMP2  & \quad  \quad NOCCD& \quad  \quad  \quad  cTPSS  &\quad  \quad cTPSSh &  \quad  \quad TPSSrsh  \\
\vspace{-0.2cm}\\
 \multicolumn{9}{c}{Free orientation}\\
\vspace{-0.2cm}\\
4.2553191e-06 & 2.9039E-13  & 3.8047E-13  & 3.7255E-13  & 4.0547E-13  & 3.6818E-13  & 4.3102E-13  & 3.2497E-13  & 2.1987E-13  \\
0.0042553191 & 2.9023E-10  & 3.8024E-10  & 3.7227E-10  & 4.0539E-10  & 3.6806E-10  & 4.3087E-10  & 3.2477E-10  & 2.1977E-10  \\
0.0127659574 & 8.6773E-10  & 1.1359E-09  & 1.1114E-09  & 1.2140E-09  & 1.1013E-09  & 1.2889E-09  & 9.6976E-10  & 6.6076E-10  \\
0.0212765957 & 1.4392E-09  & 1.8818E-09  & 1.8413E-09  & 2.0156E-09  & 1.8268E-09  & 2.1231E-09  & 1.6096E-09  & 1.1034E-09  \\
0.029787234 & 2.0061E-09  & 2.6185E-09  & 2.5641E-09  & 2.8039E-09  & 2.5412E-09  & 2.9383E-09  & 2.2396E-09  & 1.5437E-09  \\
0.0382978723 & 2.5683E-09  & 3.3432E-09  & 3.2753E-09  & 3.5730E-09  & 3.2431E-09  & 3.7764E-09  & 2.8571E-09  & 1.9781E-09  \\
0.0468085106 & 3.1206E-09  & 4.0484E-09  & 3.9676E-09  & 4.3202E-09  & 3.9299E-09  & 4.5638E-09  & 3.4600E-09  & 2.4065E-09  \\
0.0553191489 & 3.6593E-09  & 4.7297E-09  & 4.6391E-09  & 5.0450E-09  & 4.5973E-09  & 5.3258E-09  & 4.0465E-09  & 2.8311E-09  \\
0.0638297872 & 4.1854E-09  & 5.3867E-09  & 5.2893E-09  & 5.7431E-09  & 5.2424E-09  & 6.0562E-09  & 4.6157E-09  & 3.2520E-09  \\
0.0723404255 & 4.7004E-09  & 6.0196E-09  & 5.9168E-09  & 6.4113E-09  & 5.8648E-09  & 6.7503E-09  & 5.1669E-09  & 3.6681E-09  \\
0.085106383 & 5.4483E-09  & 6.9180E-09  & 6.8123E-09  & 7.3561E-09  & 6.7544E-09  & 7.7215E-09  & 5.9632E-09  & 4.2837E-09  \\
0.0957446809 & 6.0463E-09  & 7.6175E-09  & 7.5126E-09  & 8.0876E-09  & 7.4518E-09  & 8.4727E-09  & 6.5881E-09  & 4.7893E-09  \\
0.1063829787 & 6.6224E-09  & 8.2727E-09  & 8.1714E-09  & 8.7742E-09  & 8.1119E-09  & 9.1809E-09  & 7.1868E-09  & 5.2836E-09  \\
0.1276595745 & 7.7287E-09  & 9.4717E-09  & 9.3930E-09  & 1.0022E-08  & 9.3227E-09  & 1.0478E-08  & 8.3394E-09  & 6.2514E-09  \\
0.1489361702 & 8.7708E-09  & 1.0532E-08  & 1.0479E-08  & 1.1125E-08  & 1.0423E-08  & 1.1582E-08  & 9.3909E-09  & 7.2405E-09  \\
0.170212766 & 9.8030E-09  & 1.1515E-08  & 1.1497E-08  & 1.2134E-08  & 1.1449E-08  & 1.2579E-08  & 1.0386E-08  & 8.1966E-09  \\
0.1914893617 & 1.0821E-08  & 1.2439E-08  & 1.2458E-08  & 1.3107E-08  & 1.2421E-08  & 1.3483E-08  & 1.1359E-08  & 9.1868E-09  \\
0.2127659574 & 1.1893E-08  & 1.3375E-08  & 1.3438E-08  & 1.4109E-08  & 1.3421E-08  & 1.4340E-08  & 1.2288E-08  & 1.0239E-08  \\
0.2340425532 & 1.3027E-08  & 1.4353E-08  & 1.4453E-08  & 1.5179E-08  & 1.4462E-08  & 1.4974E-08  & 1.3184E-08  & 1.1228E-08  \\
\vspace{-0.2cm}\\
 \multicolumn{9}{c}{Fixed orientation}\\
\vspace{-0.2cm}\\
4.2553191e-06 & 1.9322E-13  & 2.9405E-13  & 2.8695E-13  & 3.2777E-13  & 2.8460E-13  & 3.3483E-13  & 2.2953E-13  & 1.2543E-13  \\
0.0042553191 & 1.9315E-10  & 2.9392E-10  & 2.8682E-10  & 3.2472E-10  & 2.8200E-10  & 3.3468E-10  & 2.2949E-10  & 1.2542E-10  \\
0.0127659574 & 5.7771E-10  & 8.7849E-10  & 8.5749E-10  & 9.2936E-10  & 8.4313E-10  & 1.0004E-09  & 6.8673E-10  & 3.7565E-10  \\
0.0212765957 & 9.5749E-10  & 1.4539E-09  & 1.4203E-09  & 1.5203E-09  & 1.3968E-09  & 1.6553E-09  & 1.1387E-09  & 6.2398E-10  \\
0.029787234 & 1.3304E-09  & 2.0159E-09  & 1.9713E-09  & 2.1161E-09  & 1.9388E-09  & 2.2923E-09  & 1.5814E-09  & 8.6909E-10  \\
0.0382978723 & 1.6938E-09  & 2.5604E-09  & 2.5060E-09  & 2.6825E-09  & 2.4649E-09  & 2.9046E-09  & 2.0106E-09  & 1.1097E-09  \\
0.0468085106 & 2.0448E-09  & 3.0828E-09  & 3.0203E-09  & 3.2354E-09  & 2.9708E-09  & 3.4886E-09  & 2.4231E-09  & 1.3449E-09  \\
0.0553191489 & 2.3821E-09  & 3.5791E-09  & 3.5106E-09  & 3.7505E-09  & 3.4531E-09  & 4.0408E-09  & 2.8156E-09  & 1.5736E-09  \\
0.0638297872 & 2.7039E-09  & 4.0467E-09  & 3.9756E-09  & 4.2310E-09  & 3.9103E-09  & 4.5536E-09  & 3.1866E-09  & 1.7949E-09  \\
0.0723404255 & 3.0091E-09  & 4.4837E-09  & 4.4118E-09  & 4.6751E-09  & 4.3389E-09  & 5.0242E-09  & 3.5367E-09  & 2.0088E-09  \\
0.085106383 & 3.4351E-09  & 5.0769E-09  & 5.0100E-09  & 5.2639E-09  & 4.9273E-09  & 5.6533E-09  & 4.0150E-09  & 2.3147E-09  \\
0.0957446809 & 3.7595E-09  & 5.5143E-09  & 5.4555E-09  & 5.7052E-09  & 5.3657E-09  & 6.1054E-09  & 4.3717E-09  & 2.5551E-09  \\
0.1063829787 & 4.0577E-09  & 5.9013E-09  & 5.8542E-09  & 6.0770E-09  & 5.7579E-09  & 6.4941E-09  & 4.6911E-09  & 2.7828E-09  \\
0.1276595745 & 4.5810E-09  & 6.5380E-09  & 6.5210E-09  & 6.6937E-09  & 6.4137E-09  & 7.1006E-09  & 5.2277E-09  & 3.2042E-09  \\
0.1489361702 & 5.0254E-09  & 7.0221E-09  & 7.0408E-09  & 7.0268E-09  & 6.9254E-09  & 7.5196E-09  & 5.6535E-09  & 3.5855E-09  \\
0.170212766 & 5.4161E-09  & 7.4009E-09  & 7.4569E-09  & 7.2651E-09  & 7.3360E-09  & 7.8146E-09  & 6.0030E-09  & 3.9401E-09  \\
0.1914893617 & 5.7786E-09  & 7.7247E-09  & 7.8115E-09  & 7.6384E-09  & 7.6874E-09  & 8.0475E-09  & 6.3133E-09  & 4.2836E-09  \\
0.2127659574 & 6.1282E-09  & 8.0366E-09  & 8.1425E-09  & 7.8679E-09  & 8.0169E-09  & 8.2742E-09  & 6.6080E-09  & 4.6215E-09  \\
0.2340425532 & 6.4749E-09  & 8.3608E-09  & 8.4754E-09  & 8.0490E-09  & 8.3533E-09  & 8.5219E-09  & 6.9088E-09  & 4.9554E-09  \\
\hline
\end{tabular}    
\end{center}
\end{table}